# Superconductor relaxation - A must to be integrated into stability calculations

Harald Reiss

*Department of Physics*
*University of Wuerzburg, Am Hubland, D-97074 Wuerzburg, FRG*
harald.reiss@physik.uni-wuerzburg.de

Abstract

A superconductor is stable if it does not quench. Quench is a short-time physics problem. For its deeper understanding of, and how to avoid quench, the physics behind stability has to be analysed. A previously suggested dynamic relaxation model is re-considered and applied to YBaCuO 123 and BSCCO 2223 high-temperature, thin film superconductors. Parallel to this investigation, an unconventional approach using an electrical resistance network (a cell model) is applied to introduce a method how to estimate the extent by which, in resistance measurements, exact determination of critical temperature of superconductors is possible. This resistive cell model, when considering its numerical convergence behaviour, in a side result may provide an alternative explanation of (at least a contribution to) bending of resistivity vs. temperature curves, and perhaps also an alternative to standard explanations of the thermal fluctuations impact on these curves.

The dynamic relaxation and the resistance models provide a parenthesis that correlates, in terms of the Ginzburg-Landau order parameter, (i) solution of superconductor stability problem (the main objective of this paper), with tentative explanation of (ii) bending of the resistivity curves near critical temperature and (iii) with predictions from thermal fluctuations.

Keywords

Superconductor phase transition; relaxation; critical temperature; resistive cell model; multi-component heat transfer, convergence

# 1 Superconductor stability

A superconductor is stable if it does not quench. Quench can be avoided by application of stability models to design, manufacture, handling and operation of superconductors.

Traditional stability models are described in [1 - 3]. With only few exceptions, stationary states are standard conditions in traditional, stationary stability calculations.

Quench is a short-time physics problem. As extension of standard stability calculations, numerical methods have been suggested by the present author [4 - 6]. The results improve reliability of stability predictions by local results of temperature fields and critical and transport current density, especially in case of situations close to phase transition. But, and this is the main subject of this paper, superconductor stability against quench, with numerical simulations, still is not at its end. A new holistic, multi-physics approach is needed to treat the stability problem adequately. It requests integration of the relaxation problem into the stability analysis.

A recently suggested "microscopic stability model" [7] has provided the core of the numerical stability calculations (this model essentially is a relaxation model, see later). Subsequent stability papers have refined this model, from filaments to thin film investigations [8].

Stability analysis needs taking into account relaxation of the superconductor after disturbances. The property "microscopic" means that calculation of relaxation time of the electron system, after a disturbance, is performed by step by step particle-particle selections (single electrons as candidates for pair re-organisation). In the present paper, the model is integrated into (is a correction to) the proper stability calculations that as previously are performed numerically to yield temperature vs. time excursion; these are provided by extensive Finite Element (FE) simulations. From the FE results, the relaxation model yields corresponding, final equilibrium temperatures (the temperature of the state when relaxation is completed). The question is after which experimental or simulation time relaxation will be accomplished.

As a general rule, thermal equilibrium is obtained in a superconductor substance when it fulfils the Meissner effect (a state that is independent of the history, i. e. regardless whether it is obtained under field or zero field cooling conditions). This is accounted for in the numerical calculations in each of the very large number of single FE elements.

Parallel to this investigation, an unconventional approach using an electrical resistance network (a cell model) is applied to introduce a method how to estimate the extent by which, in resistance measurements, exact determination of critical temperature of superconductors is possible.

When the stability model is integrated into the cell model, it suggests, as a side result, postulation of a non-local, transition boundary layer, expressed as a temperature uncertainty, $\vartheta T_\rho$, around critical temperature. Within $\vartheta T_\rho$, the resistivity curve, under heat-up, continuously approaches the normal conduction value. Curved resistivity vs. temperature, instead of a sharp jump, is frequently observed in experiments.

As another side result, this study may also, though weakly, contribute an alternative to predictions of the thermal fluctuations problem. More importantly, the transition boundary layer, $\vartheta T_\rho$, questions existence of a sharp, uniquely defined critical temperature but allows a sharp estimate of its uncertainty when it is obtained from resistance measurements.

The dynamic relaxation model is a parenthesis that correlates, in terms of the Ginzburg-Landau order parameter, the solution of superconductor stability problem (the main objective of this paper) with side results of this study (a tentative explanation of bending of the resistivity curves near critical temperature and an alternative to standard assumptions of thermal fluctuations).

## 2 Relaxation

The literature reports magnetic relaxation measurements to determine pinning potentials, like in $Nb_3Sn$, in the BSCCO family and in $MgB_2$/Fe superconductors. In contrast to these studies, focus of the stability model [7] is on relaxation of the electron system from *disturbances* like transient local heat sources, usually the origin of a quench.

We do not address *statistical* electron pair decay and recombination in dynamical equilibrium. These relaxations, at any temperature $0 \leq T < T_{Crit}$, exist without disturbances. Instead, the paper is focused on electron pair decay and re-condensation, after temporal, local disturbances, of the decay products to electron pairs.

A second question is whether relaxation rates after disturbances, possibly under thermal run-away, are large enough to re-organize superconductor stability to obtain a new thermodynamic equilibrium, as the *sole* state by which zero-loss current transport is possible.

Superconductor stability against quench thus is an aspect of the relaxation problem.

Analysis of relaxation requests calculation of relaxation time and relaxation rates. Critical current density after a disturbance neither is uniform in the conductor cross section nor is it constant in time, compare [4 - 6]. Accordingly, relaxation time and relaxation rate, too, would not be uniform and not be constant in time, if temperature distribution is not uniform but highly diversified and if it is transient, a situation that invariably happens after local disturbances.

## 2.1 A recapitulation

Excitations (thermal, magnetic, or simply from transport current exceeding critical current density, the flux flow state) disturb dynamic equilibrium between

(i) decay of electron pairs and
(ii) single electron (and quasi-particle) recombination rates.

In the equilibrium state, both decay and recombination rates are equal. In a vivid description, the meaning of quasi-particles is explained by R. D. Mattuck (see Appendix, after Eq. 11).

In the literature, relaxation of the superconductor electron system from an excited state to a new dynamic equilibrium, is expected to follow an exponential decay law, exp(-t/τ), of the density, $c(\mathbf{x},t)$, of excited electron states, by recombination to electron pairs. The relaxation time, τ, usually is considered as constant.

But relaxation cannot be completed instantaneously. It is more realistic to calculate relaxation time as a sensitively *time-dependent* quantity, τ = τ[T(t)], see later. This step has been realized in the relaxation model [7], by a multi-physics approach using analogies between relaxation in superconductors with re-organisation processes in nuclear and multi-particle physics.

The question is not whether the decay products might be correlated or uncorrelated (in the uncorrelated case they might not find partners in the strict BSC sense: no electron with momentum -**k'** and spin down, no electron with momentum -**k''** and spin up might be available [27], p. 178 - 180). Calculations of the relaxation time, as performed in [7], primarily need *distances* over which correlations between appropriate partners can be initiated, under observation of selection rules.

In analogy to Sects. 2 to 5 of [7], we consider, after a disturbance, the decay of the concentration, c(**x**,t), of decay products as

dc(**x**,t)/dt = (∂c(**x**,t)/∂x) (∂x/∂t) + ∂c/∂t                                        (1)

with two contributions: "decay in space" = (∂c(**x**,t)/∂x) (∂x/∂t) and "decay in time" = ∂c(**x**,t)/∂t.

Because of propagation of the thermal wave initiated by a disturbance, decay *in space* means that the increased concentration, c(**x'**,$t_1$ > $t_0$), resulting from a disturbance occurring at any arbitrary position, **x'**, of excited states, is distributed by a transport process to positions **x** ≠ **x**'. A transport mechanism that particularly simply can be described is diffusion, but it has to be shown that diffusion, while it is a simply described transport mechanism, is the realistic assumption, see later (Appendix).

Decay *in time* means that the disturbed, total wave function, $\psi(\mathbf{x}, t > t_0)$, that describes all electron states and decay products, returns to its equilibrium shape by recombination processes at al co-ordinates, **x**.

The steps "decay in space" and" decay in time" in Eq. (1) have extensively been described previously [7] and will not be repeated here. The first term of this equation has been shown to be small against the second.

Contributions to relaxation time from both items have to be summed up to lifetime, $\tau = t_{Eq}$, of the *total* disturbed system until dynamic equilibrium of the *total* electron state is accomplished; time $t_{Eq}(T)$ thus is the time interval that the system, at all co-ordinates, **x**, needs to arrive at the new equilibrium, at the same temperature, T', that the system attains under a disturbance above the smaller T of the preceding step.

Calculation of total lifetime, $\tau$, has to strictly follow selection rules, using "Coefficients of fractional parentage" (cfps), a concept applied in atomic and nuclear physics, the handling of which has to obey the Pauli exclusion principle.

Because of the Pauli principle, calculation of, $\tau$, has to proceed in a *step-wise* manner, with calculation of the cfps in each step, at each temperature, T' (a "sequential model"). Calculation of τ, accordingly proceeds in a series of temperature steps the individual values of which are provided as results from the FE calculations of temperature excursions. Calculation of τ is the more complex the larger the temperature and, as a consequence, the larger the number of decayed electron pairs. Their number and, accordingly, the number of calculations steps, depends on the density of electrons at given temperature. The number of electron pairs that may result from recombination of single electrons to pairs is given by the Ginzburg-Landau order parameter.

In their theory of phase transitions, Ginzburg and Landau postulated existence of an order parameter, Ψ, that in an expansion of the free energy density,

$$g_S = g_N + a(T) |\Psi|^2 + 1/2\, b(T) |\Psi|^4 + ... \qquad (2a)$$

*within the active part of the electrons,* determines the number of electron pairs available for zero-loss current transport. (for explanation of the functions a(T) and b(T) in Eq. (2a), see standard volumes on superconductivity, e. g. [9], p 72). Eq. (2a) applies to zero magnetic field. In its *original* formulation, Ψ was assumed as an unspecified physical quantity (a complex number) that characterizes the thermodynamic state of the superconductor,

$$\Psi = 0 \text{ for } T > T_{Crit}, \text{ while otherwise } \Psi = \Psi(T) \qquad (2b)$$

The square $|\Psi|^2$ is identified as the density of electron pairs. In the present paper, we consider the ratio

$$f_S = n_S(T)/n_S(T=0) \tag{2c}$$

of the density of electron pairs, $n_S$, to characterize the state of the superconductor, at temperature $T < T_{Crit}$, in relation to its value at $T = 0$. Apart from the constant $n_S(T=0)$, $f_S$ in Eq. (2c) equals $|\Psi|^2$. The $f_S$ fulfil, like in the original formulation of the order parameter, the conditions in Eq. (2b).

In the following, the ratio $f_S = n_S(T)/n_S(T=0)$ therefore is assigned "the order parameter" (it is a direct approach to the state, $\Psi$, of the superconductor by the rather simple ratio $f_S$, Eq. 2c). Complications thus do not arise when relaxation time, $\tau$, has to be calculated. This is the proper subject of [7], i. e. calculation of relaxation time; in this reference, we do not determine wave functions attained by the system at intermediate reorganisation steps.

Relaxation does not mean recombination of a *strictly limited number* of single electrons (the decay products from a disturbance) to electron pairs but re-organisation of the *total* electron body, i. e. of *all* electrons as far as they are "available".

Available means: The percentage of electrons, the active ("busy") part, $\xi$, of the total electron body, that thermally can be excited to states located in small energy intervals above $E_F + \Delta E$ and below $E_F - \Delta E$ and that contribute by

(i) electron pairs, in dynamic equilibrium with
(ii) single electrons

to thermal and electrical transport and also constitutes the electron contribution to total specific heat (and is responsible for the Meissner effect).

In the literature, the fraction, ξ, is indicated, for low-$T_C$ and high-$T_C$ superconductors, as roughly 0.1 and 10 per cent of the total electron body, respectively. It is not clear the ξ = 10 per cent contribution in high temperature superconductors might be very precise, rather it is just a rough estimate.

Using ΔE = 60 meV at T = 0, with the classical, BCS T-dependence of ΔE and with 2 ΔE(T = 0) = 3.52 $k_B T_{Crit}$ and $k_B$ the Boltzmann constant, according to [10], p. 355, the numerical constant in this relation is between 5 and 7 for high temperature superconductors (HTSC) instead of 3.52. With $E_F$ = 1 eV from [11] or from ARPES measurements, we have, at temperature not very close to $T_{Crit}$, e. g. ξ(T = 80 K), a value ξ of below 1 per cent. With increasing temperature, ξ decreases, since the number of electron pairs is successively thinned out.

In view of these uncertainties, we can only perform a sensitivity analysis showing the impact by which the ratio, ξ, indicating the active part in relation to the total electron body, is reflected by the equilibrium temperature excursions (see later, Figure 7c).

## 2.2  Modelling alternatives – A critique

Contrary to the *stepwise*, cfp-organized procedure [7, 8], an analytic, continuum expression for the electron pair density can be found in Eq. (8) of [12], with $n_S(T)/n_0$ = 1 - $(T/T_{Crit})^4$ and $n_0$ the total number of (single?) electrons at T = 0.

This expression neglects the dynamic aspects of the relaxation process. It does not explicitly integrate the step-wise sequence of a large number of statistical, single electron or quasi-particle generation and recombination processes. Instead, Eq. (8) in [12] implicitly assumes *simultaneous* re-organisation of the decay products to a new, recombined set of electron pairs and single particles; it does not consider this process as a *sequence* to restore the whole, active plus inactive parts of the total electron body to a new dynamic equilibrium. Note that this equation considers the ratio of electron *pair* density at a temperature, T, to the density, $n_0$, of *single* electrons at T = 0. It is not clear how in [12] the density $n_S(T)$ was derived from $n_0$.

A correct calculation of the re-organisation of decay products to electron pairs has to include, for each electron, n, that "looks" for (selects) its partner, n', to form a pair, the calculation of its *own* cfps from the *previous* N - 1 completed re-organisations. This means it requests expanding the wave function of a new, completed recombination state (N) (from the wave functions composed of single particle states) in terms of

(j) each of the foregoing expansions of the (N - 1)-states and
(jj) incorporating the wave function of the new pair (n, n') into the new state, N.

Each of the steps (j), (jj), has to observe the Pauli selection rule. Since it implicitly assumes *simultaneous* re-organisation of the decay products, the concept suggested in [12] violates this principle.

Fortunately, both the direct (stepwise, microscopic) method [7]) and, with some additional assumptions, the apparently heuristic Eq. (8) in [12], at least allows approximate calculation of relaxation rates and relaxation time.

Relaxation *rates* in *both* concepts [7] and [12] converge to zero when the system during warm-up very closely approaches its super-conduction/normal conduction phase transition (see later, Figure 3). Relaxation *time*, as a consequence of decreasing relaxation rate, as is shown in [7] and, later, in Figure 11 of [24], Part B, increases the more the closer the electron system during warm-up approaches this phase transition.

This conclusion contradicts Buckel and Kleiner [13], Chap. 4, p. 262. The authors state that in conventional superconductors the probability that an unpaired electron finds a suitable partner for recombination to form a electron pair *decreases* under increasing temperature. But the present situation, relaxation from disturbances, is strongly different. This can be seen as follows (we have to look very carefully onto this situation):

Starting from an original, dynamic equilibrium at a temperature, T, followed by a disturbance at this temperature (under continued warm-up leading to a temperature T' > T), increasingly *more* single particles, have to recombined to pairs in order to generate, by reorganisation of the *total* electron body, a new dynamic equilibrium at T'. Statistically, more "partners" to form a pair become available, simply because more pairs decay if temperature increases. This statistical, un-displaced dynamic equilibrium, as one step out of a great number of analogue ones, like all others can be obtained only if relaxation has been *completed* at each of the (intermediate) foregoing temperatures, T'.

The increased number, n(T') > n(T), of single electrons *within* the active body that has to be reorganised to pairs, requests a *larger* number n(T') of single electrons (the decay products) to identify partners, n'(T'), under observation of the said selection rules (cfps and the Pauli principle), which means the whole recombination process takes more time the more the system approaches the phase transition.

Another potentially alternative explanation is suggested in the papers by Gray et al. [14,15]. But they do not refer to a temperature increase from absorption of a radiation pulse or from other thermal disturbances. Contrary to the situation described in [7], Gray's papers describe *injection* experiments to create an *additional* number, Δn, of quasi-particles, a strongly non-equilibrium state *above* the "core" ($N_{Total}$) of bound electrons in the undisturbed superconductor. This model apparently has not taken into account the cfp-rule that must be observed *within* the $N_{Total}$(T') active state. Injection experiments, as a consequence, cannot be described by the rigorous, sequential model [7].

The model [7] therefore contradicts [13] (but is not contradicted by this reference), because the balance to yield N in [13] is not complete. The model [7] also contradicts [14, 15] (but is not contradicted by injection experiments) because in [7] additional electrons are not delivered to the previous state.

The conclusion from [7] also specifies a note that Annett [9], added on p. 52 to the left column (and again requests careful inspection). There we have: "The word superconductor is used only to mean a material with a *definite* phase transition and critical temperature." (correction by the present author: "definite", here written in Italics, probably means "completed"). But critical temperature is a dynamic, *equilibrium* electron state unto which a series of *non-equilibrium* states may converge when *no more disturbances*, besides statistical equilibrium fluctuations between decay of electron pairs and re-condensation of decay products to pairs, have to be compensated.

Another item to be explained is determination the spatial structure of electron pairs in a superconductor volume.

It is hardly correct to assume, as has been made in [16], that the electron pairs occupy a specific superconductor volume in that they fill these volumes uniformly (and that magnetic field lines would penetrate this volume in-between neighbouring, possibly overlapping electron wave functions).

In the present paper, electron pairs are understood as *correlations*, in a correlated Fermi system a *statistical* concept, they are not solid particles that would densely occupy a superconductor volume. Complications (like penetration of magnetic field lines, mentioned in [16]) are avoided (there are no collisions between field lines and piles of overlapping electron pairs), and there are no impacts on centre of mass motion, spin of the electron pair or, in particular, variable distance of the two electrons that constitute the pair.

A more attractive explanation of the spatial structure of electron pairs, directly from BCS-theory, is provided in [17].

As mentioned, electron pairs cannot be regarded as minute, *materials* entities (a solid, coherent particle of charge 2e). Instead, they are *correlations* between electrons j and k, with the j and k *continuously exchanged* (replaced by other electrons j' and k' that randomly, but under observation of quantum-mechanical selection rules, are identified from a very large number (a statistically filled *repository* within the active electron body), of single electrons or quasi-particles. The selection and recombination processes, replacements of j and k by j' and k', as simulated in [7] all are statistical processes.

Yet the explanation [17] of the spatial structure of electron pairs will tentatively be applied in Sect. 3 to derive a "pseudo-porosity" from a cell model, a resistive network of the superconductor transport properties, as a function of the order parameter (like standard porosity a continuum property of a solid or liquid).

In dynamic equilibrium, the porosity separates the number of electron pairs from the number of single electron or quasi-particle states (strictly speaking, their densities).

Another problem arises with respect to current across a superconductor/normal conductor interface: Can it be described always as a "transport process"? Which conditions get electrical current or heat flow a "transport" process? This is not very trivial, but the discussion is postponed to the Appendix where some general aspects of transport processes (electrical and, as a parallel, multi-component, heat transfer processes) are discussed.

# 3 Curvature of resistivity vs. temperature

The literature reports a great number of electrical resistivity, $\rho_{El}(T)$, or of conductivity, $\sigma_{El}(T)$, measurements that yield curves with remarkable deviations from the expected, sharp increase of $\rho_{El}(T)$ or of decrease of $\sigma_{El}(T)$ at $T_{Crit}$. This is schematically illustrated in Figure 1.

We make an attempt to explain, by a two-component, resistive cell model, the observed rounding of the transition curves by a correlation with variations of the superconductor order parameter (here in the approximation suggested in [7], compare Eq. 2a-c of the present paper).

Glover III [18] claims evidence that the observed rounding of $\sigma_{El}(T)$, the "thermal fluctuations problem", in thin samples with short electron mean free paths relies on variations of an intrinsic materials property, a "sort of fluctuating superconductivity".

The cell model described below does not rely on *fluctuations* of superconductivity (in [18], it is not very clear what is really meant by fluctuations against/between what, and to which extent).

The cell model instead relies on calculation of an effective resistivity or conductivity of *two phases* co-existing in parallel, not separated in space, in a *common* superconductor material. The model considers a sole (electron) body that according to its temperature separates into two thermodynamic, different *phases*, not into different *materials* and accordingly is not identical to e. g. two-liquid models.

Two-fluid models are well known in standard superfluidity and superconductivity literature, in classical fluid dynamics and in nuclear physics where a liquid-drop model (the nucleus imagined as an incompressible liquid) serves for calculation of nuclear binding energy.

In contrast to [18], there is, in the present model, no fluctuation *of one* but variable contributions by *two* different, thermodynamic phases within the *same* material. Under dynamical equilibrium, their contribution to effective electrical resistivity relies solely on a function of the order parameter (the approximation given in Eq. 2c), that is a unique, strong function of temperature. Temperature of course may vary locally, under e. g. non-uniform materials properties or from locally different transport to critical current density or from locally different magnetic fields.

The two-phase model in the following applies a resistance network. Description of the effective resistivity of this network is given in terms of a (pseudo-) porosity. This has to be defined clearly. The porosity is solely based on the order parameter of the superconductor electron system and its temperature dependency. It separates the density of single electrons and quasi-particles from the density of electron pairs.

### 3.1 Cell model and resistance network

The Russell cell model [19] is a standard tool for application to two-component systems to describe their effective materials and transport properties. It was originally derived for solely conductive, thermal energy transport but applies to also electrical, in general to any conductive transport process provided it

(i) is really a "transport" process and, if so,
(ii) is expressed by a differential formalism (as is done in Fourier's differential equation in case of solely conductive heat transfer).

Network and cell model will be applied first to the region $T < T_{Crit}$. A tentative, but well-explained extension to $T > T_{Crit}$ will be made in Sects. 4 and 5).

Let $k_{Core}$ denote, in a traditional application of this model, the thermal conductivity of spherical "inclusions" completely embedded in a continuum (a matrix or "shell"). In a thermal super-insulation, for example, this is the vacuum, or a gas of at least a very low residual pressure.

In a continuous thermal super-insulation, the inclusions (the "core") in the evacuated space are minute, solid particles (powders or fibres; we do not consider evacuated, highly reflecting multi-foils). The thermal conductivity (if it *exists*, see again the Appendix, or compare [25, 26]) of the surrounding continuum (the "shell", here the vacuum) relies on residual gas pressure. If $k_{Shell}$ denotes thermal conductivity of the shell, the ratio $k_{Core}/k_{Shell}$ in thermal super-insulations is very large.

The procedure is applicable to any other continuum, not only to evacuated space filled with solid particles.

If porosity, $\Pi$, is known, the effective, integral conductivity, k, according to the Russell cell model reads

$$k = k_{Core} [\Pi^{2/3} + (k_{Core}/k_{Shell}) (1 - \Pi^{2/3})]/[\Pi^{2/3} - \Pi + (k_{Core}/k_{Shell}) (1 - \Pi^{2/3} + \Pi)] \qquad (3)$$

and applies to also resistivity when k is replaced by $\rho = 1/k$.

In order to apply the Russell cell model, or any other traditional cell model, to also *superconductors*, the "inclusions" (the core) have to be interpreted as being composed of sets of electron pairs, and the "shell" is given by the overwhelming number of single electrons or quasi-particles, all of them as decay products arising from disturbances.

To get application of Eq. (3) working also in case of superconductors, we have to define the corresponding "pseudo-porosity", Π.

With increasing temperature, the number of "inclusions" (sets of electron pairs) goes to zero at the phase transition; the porosity of the total, then normal conducting (NC) electron body accordingly approaches Π = 1. The electrical resistivity of this state, a finite value, $\rho_{El,NC}(T)$, is obtained from experiments.

Electron pair, *zero* resistivity, $\rho_{El,Core} = \rho_{El,SC}(T)$, cannot successfully be applied in Eq. (3). It has to be represented by a finite, *non-zero* resistivity, $\rho_{El,Core} > 0$. This (pseudo-) resistivity of the superconducting, zero-resistive transport channel, if it is a transport process at all (if it "exists"), must be smaller, by many orders of magnitude (at least 20), than the resistivity of the normal conducting phase, the shell of single electrons or quasi-particles (the decay products).

The ratio 20 is not the real problem; instead, it is the "*existence*" of resistivity or conductivity components. This situation shall be explained by an analogy of this problem known from multi-component heat transfer. There, it is the existence of total thermal conductivity or thermal resistivity that is in question. In case any of several heat transfer components, like radiation, cannot be described as a conductivity (which may happen in thin, partly transparent films), it is not clear that total heat flow could be explained in terms of a total resistivity or conductivity, which means they are not given ("might not exist") as quantifiable expressions). See again the Appendix for more explanation.

Predictions of the cell model may diverge at very small and very large values of the porosity (this is apparently a weak point of all cell models). Application of this model to superconductors thus might become critical when, under warm-up, the number of electron pairs near $T_{Crit}$ goes to zero when, as mentioned, the porosity approaches $\Pi = 1$.

In the present case, the resistivity must converge to the (as assumed, very small) finite value $\rho_{El,Core}$ if $\Pi \to 0$. If this is the case, the cell model is confirmed to be applicable to the present problem, at least in view of this condition. Formally, satisfaction of this and other single conditions can easily be checked (here, just by assuming $\Pi = 0$ in Eq. 3).

A more complete, general check whether the procedure works correctly requests a set 1 to 7 conditions to be fulfilled, see the contribution [20], p. Deb 1, and literature cited therein, how to calculate conductivity of particle beds. When conductivity is translated into resistivity, and (for example, condition 2) $\rho_{Bed} = \rho_{El,eff} \to \rho_{El,Shell}(T)$, for $\Pi = 1$, all these conditions (with the exception of No. 7 that does not apply), are satisfied by the procedure described in the present paper (an Excel sheet). But it is necessary, and has to be controlled, too, that also all elements of a *numerically generated set* of resistivity, $\rho_{El}(T)$, in dependence of *temperature*, not only a *single* value, converge to the limits given in conditions 1 to 6 of [20], compare Figure 6b.

For Figure 2b, a finite non-zero resistance originating from weak links would request incorporation of another rectangle, but is neglected in this presentation (its contribution shall be enclosed in the blue rectangles). In continuous thermal super-insulations, weak links result from e. g. contact resistances between neighbouring solid particles, in superconductivity, weak links may arise from a variety of local resistances, in high-temperature superconductors between neighbouring grains (like Josephson barriers to electrical transport, contact angle mis-orientations).

### 3.2 Calculation of the pseudo-porosity

Porosity in the following has to be explained as a function of the ratios by which electrons, in their two conducting phases, contribute to total electrical conductivity. Contribution of the phases is solely related to temperature dependence of the ratio of particle (electron pair) density at temperatures $T$ and $T_0$ ($T_0$ taken as a reference value). At different temperature, this ratio, $f_S(T) = n_S(T)/n_S(T_0)$, as mentioned approximates the Ginzburg-Landau order parameter (strictly speaking, its square),

Using $n_S$ for density of electron pairs, and since $n_S$ depends strongly on temperature, the fraction $1 - f_S$ needed for derivation of the (pseudo-) porosity cannot be constant, which means the pseudo-porosity, too, is a function of temperature, $\Pi(T) = 1 - f_S(T)$, and since temperature depends on time, $T = T(t)$, under disturbances, also the porosity is a function of time during a disturbance.

The number, $N_{CP}$, of electron pairs in an arbitrarily, small volume, $V_{SC}$, of the superconductor material then is given by

$$N_{CP} = f_S(T)\, n_S(T_0)\, V_{SC} \qquad (4)$$

The distance of two electrons forming an electron pair can be approximated roughly by the Pippard coherence length ξ (the range of the wave function for relative motion of the electron pair, Blatt [27], p. 161-162) or (see below), when including its temperature dependency, by the length ζ(T) in Eq. (7). In both BSCCO 2223 and YBaCuO 123 superconductors, ξ amounts to about 10 nm; in NbTi, ξ taken from [13] is 5.8 nm.

Spatial size of one pair (as mentioned, not a materials volume, not a solid particle of charge 2e but simply a correlation between two real particles, each of charge 1e), within which *correlation* of the two electrons exists, is $dV_{CP}$. With the Pippard correlation length, we can estimate a corresponding (Pippard) correlation *volume*, which again means: not a materials volume. The total *correlation* volume, $V_{CP}$, of *all* electron pairs $N_{CP}$ in the volume $V_{SC}$, in the rough approximation, then results from $V_{CP} = N_{CP}\, dV_{CP}$.

In comparison to the average distance of single electrons (in the order of 0.1 nm), the large (internal) size of a pair covers a very large number of electrons, which results in strong overlap of correlation integrals between corresponding wave functions of different pairs.

A Cooper pair of size (correlation length) $10^2$ to $10^3$ nm, with atoms of size $10^{-1}$ nm covers $(10^3)^3$ to $(10^4)^3$ electrons which means the correlation of each Cooper pair overlaps with "billions and billions of others, very different from a Bose gas" [21]. Thus the effective correlation volume (in this probably oversimplified picture), and with M the number of overlaps, reads

$$V_{CP,eff} = (N_{CP}\, dV_{CP})/M \tag{5}$$

from which the porosity

$$\Pi = 1 - V_{CP,eff}/V_{SC} \tag{6}$$

The number M of overlapping electron pairs cannot be constant but is a function of temperature, T. This follows from the temperature dependence of the coherence length in the dirty limit,

$$\zeta(T) = 0.85\, (\zeta_0\, \ell)^{1/2}\, [T_{Crit}/(T_{Crit} - T)] \tag{7}$$

see [22], p. 210, with $\zeta_0$ the Pippard coherence length (at T near zero) and $\ell$ the mean free path of electrons, respectively. The length $\zeta(T)$ can be interpreted as the mean 1D dimension of an electron pair.

With increasing T, the length $\zeta(T)$ increases, which means, with $\varsigma \approx \ell$, the mean distance between two conduction-band electrons (about 0.1 nm), and $\zeta_0$ the Pippard coherence length, about 100 to 1000 nm, the ratio $\zeta(T)/\varsigma$, and accordingly, M, becomes very large.

Eq. (6) of [17], with a square dependence $M \sim (E_F/\Delta E)^2$ should provide an improved estimate of M, instead of simple ratios $\zeta_0/\varsigma$ applied in [21] and $\zeta_0(T)/\varsigma$ using Eq. (7).

In any case, within one electron pair, there is a very large number of other electrons that in turn correlate to electron pairs and overlap within this distance, as schematically indicated in Figures 1 an 4 of [17]. The number M of overlaps in YBaCuO 123 within $V_{SC}$ to be used in Eq. (5) thus should amount to at least M = $10^6$ to $10^9$, increasing with temperature.

Reversely, with increasing T, but *decreasing* M, $V_{CP,eff}$ would increase which means the porosity would become negative provided $N_{CP}$ is constant. But with increasing temperature, T → $T_{Crit}$, $N_{CP}$ decreases, which finally (*increasing* M, decreasing $N_{CP}$) limits $V_{CP,eff}$ strongly to very small values. The porosity thus increases, Π(T) → 1, with increasing T, and the resistivity $ρ_{El}$(Π) converges, in a steady transition, to the normal conducting value, $ρ_{El}$(T) → $ρ_{El,NC}$(T), compare Figures 4, 5 and 6a,b.

But at T very close to $T_{Crit}$, handling of the procedure collides with limitations set to the applicability of Fortran routines or of Excel sheets. When declaring real and integer variables as "Real*16", the smallest possible deviation that can be simulated, T - $T_{Crit}$, is $10^{-16}$ K. This sets a limit to the porosity obtainable from Eqs. (2) to (4).

Accordingly, numerical sets of porosity, Π(T), and of resistivity, $ρ_{El}$(T), are generated, the individuals of which, at T = $T_{Crit}$ - $10^{-16}$ K, converge to Π(T) = 1 and to the pure, normal conduction resistivity, $ρ_{El}$(T) = $ρ_{El,NC}$(T).

Figure 4 shows resistivity, $ρ_{El}$(T) = $ρ_{El,SC}$(T), of YBaCuO 123 as function of porosity, and Figures 5 and 6a the resistivity of NbTi, YBacuO 123 and BSCCO 2223 vs. temperature.

The resistivity does not diverge to un-physically large values (a minor, yet favorable argument to support applicability of the cell resistance model).

Detailed inspection of the diagrams in Figure 6a confirms the deviation from a sharp jump of $\rho_{El}(T)$ at $T < T_{Crit}$. The deviations become more obvious in Figure 6b (enlarged sections of $\rho_{El}(T)$ of the NbTi, YBaCuO 123 and BSCCO 2223 resistivity diagrams) and in Figure 6c for the electrical conductivity of the BSCCO superconductor.

As is to be expected, the more temperature approaches $T_{Crit}$, the larger, trivially, is the contribution within the Russell model to $\rho_{El}(T)$ by the normal conducting phase, $\rho_{El,NC}(T)$. This means the temperature dependence of $\rho_{El,NC}(T)$, the resistivity of the normal conducting phase, too, increasingly determines the excursion with temperature of the total resistivity, $\rho_{El}(T)$.

### 3.3   Symmetry operations to complete the model to $T > T_{Crit}$

Can we extend the results from Figures 5a and 6 a-c to $T > T_{Crit}$, in an attempt to generate the curve, $\rho_{El}(T)$, over the *total* temperature range? The attempt in the following will be restricted to $T_{Crit} - 2\ K \leq T \leq T_{Crit} + 2\ K$. This result shall be transferred to Sect. 4 to suggest a tentative (as will be shown, small) *contribution* for understanding of the thermal fluctuations problem (the full curves in Figure 6a have already been obtained from the point-symmetry operations described below).

For this purpose, we divide the plane $\rho_{El,SC}$ vs. T in Figure 5a into 4 quadrants and map the curve $\rho_{El,SC}(T)$ from quadrant 4 to quadrant 1 by three, successively performed, elementary symmetry operations (Figure 5b).

First, two operations, $\rho_{El,SC}(T) \rightarrow \rho_{El,SC}(-T)$, with the vertical axis $T_{Crit}$ = const as the mirror plane, and $\rho_{El,SC}(-T) \rightarrow -\rho_{El,SC}(-T)$, with the vertical position of the mirror plane defined by the solid red circle (the maximum obtainable in the calculations of the resistivity in quadrant IV or III). These steps are followed by $-\rho_{El,SC}(-T) \rightarrow -\rho_{El,SC}(-T) + |\Delta\rho_{El,SC}(-T)|$. In total, all these operations simulate a point-symmetry operation of $\rho_{El,SC}(T)$ against the solid, red circle.

The linear transformation $-\rho_{El,SC}(-T) \rightarrow -\rho_{El,SC}(-T) + |\Delta\rho_{El,SC}(-T)|$, from quadrant II to quadrant I using the constant $|\Delta\rho_{El,SC}(-T)|$ serves to adjust the final $\rho_{El,SC}(T)$ in quadrant I to overlap with the normal resistivity of the superconductor, $\rho_{El,SC}(T)$, at $T \geq T_{Crit}$ (the solid, light-blue circles in Figure 5b).

The limit which is set to the approximation of $\rho_{El,SC}(T < T_{Crit})$ to the normal conduction $\rho_{El,NC}(T \geq T_{Crit})$ is given by the numerically obtainable maximum $\rho_{El,SC}(T < T_{Crit}) = \rho_{El,SC}(T = T')$. The maximum follows from limitations (as mentioned, by the maximum number of digits of integer and real type variables) experienced when temperature shall approach $T_{Crit}$ as close as possible. The final difference $\rho_{El,SC}(T = T') - \rho_{El,NC}(T \geq T_{Crit})$ is applied as a linear shift, $|\Delta\rho_{El,SC}(-T)|$, of the curve $\rho_{El,SC}(T)$ when it is mapped from quadrant II to quadrant I.

These symmetry operations are *justified* because they are applied to *solely* the superconductor electron system, $\rho_{El,SC}(T)$, not to the lattice with its zero electrical conductivity. The operations also are *applicable* because they are restricted to the small temperature interval within which the resistivity from zero steadily increases to large values.

From Figure 6b, a (non-local) boundary layer, $\delta T_\rho$, then can be identified within which this increase, near critical temperature, is completed.

The interval, $\delta T_\rho$, results from solely these symmetry operations, it is not determined in experiments performed in each quadrant. Width of this interval depends on the sensitivity by which $\rho_{El,SC}(T)$ decays from $\rho_{El,NC}(T)$ when temperature is slightly reduced below $T_{Crit}$. Values of the width are given in Sect. 5.

An interesting question is whether the derivation of the interval $\delta T_\rho$ could be applied to also the extent of the proximity effect, but this needs separate investigations.

## 4  Possible impacts on the Thermal Fluctuations problem

While the behaviour of the calculated $\rho_{El}(T)$ in Figure 6a-c qualitatively confirms bending of resistivity vs. temperature curves frequently seen in standard experiments, deviations of the calculated curves in relation to results of their measurements are much larger, in particular in regions indicated by ellipse I in Figure 1. Strong deviations from the expected sharp jump were observed during early decades of high temperature superconductivity development.

The reason for the deviations from sharp increase may be manifold: First, a straight, vertical line, at exactly $T = T_{Crit}$, could be observed only in case composition of the material is absolutely stoichiometric and perfectly in the clean limit (mean free path of electrons exceeding coherence length), in zero magnetic field and without weak-links. Non-stoichiometry and magnetic field influence presumably are most important sources of bending and its uncertainties. Trivially, derivation of the pseudo-porosity might be erroneous (or, frankly speaking, even the whole cell model might not be applicable at all to superconductors), but the observed convergence of the results speak in its favour. Also exact measurement of the curve $\rho_{El}(T)$ at temperature very close to $T_{Crit}$ is (potentially) extremely difficult, with possibly significant experimental errors.

As another argument speaking in favour of deviations from the sharp increase of $\rho_{El}(T)$ much stronger than predicted in Figure 6a-c, might come from Ginzburg-Landau theory. Remember that this theory is a mean-field theory that does not integrate thermal fluctuations. If they shall be taken into account, variations (fluctuations) of the free energy of the superconductor against its equilibrium value should be reflected by variations of the order parameter, $\Psi$. The system, with certain probability, in such a state, is characterized by an order parameter, $\Psi'$, that is not very different from $\Psi$.

Annett [9], p. 89 - 93, assumes the usual Boltzmann probability distribution for such fluctuations to occur. The same reference shows that thermal fluctuations, resulting from this distribution, can provide "very large contributions to heat capacity essentially diverging at critical temperature." (see Figure 4.8 of [9]). The effect might clearly be seen in high temperature superconductors, in their heat capacity and also in their $\rho_{El}(T)$-curves near $T_{Crit}$.

But in both cases [18, 9], the physical reasons for the fluctuations are not very clearly explained. Both references insist on probabilities ("statistical fluctuations of superconductivity", and "Boltzmann statistics") but do not provide a detailed explanation of the *physical* background why fluctuations within the microscopic structure of superconductors might come up at all (instead of just relating to statistics for the statistics' sake).

This remark does not question Landau's theory of second order phase transitions in superconductors. But it is striking that, as an alternative explanation, the curvature of the resisitivity vs. T diagrams, simply might result from the *T-dependence* of the order parameter. Should this observation not be mentioned?

In [23], Figure 2 (see also footnote 1 in this reference) and in our previously published papers, we indeed have applied statistical fluctuations, but these were addressed to the values of $T_{Crit}$, $B_{Crit}$ and $J_{Crit}$ as parameters in the FE calculations of transient disturbances, to account for irregularities of materials and transport properties arising during manufacture, handling and applications of the materials, which means, accounting for *really* existing, materials property or transport variations (deteriorations) observed in experiments or from information available in the literature.

This issue, strict correlation with really existing uncertainties, apparently is missing in the explanations of the thermal fluctuations problem provided in [18]. Application of the Boltzmann statistic needs justification preferentially based on physics, materials science and on experience.

In any case, bending of $\rho_{El,SC}(T)$ from the sharp increase expected at $T_{Crit}$ and deviations of the calculated $\rho_{El,SC}(T)$ from the experimental curves, can at least qualitatively be explained by the resistance model with temperature-dependent order parameter; it is thus not a "phenomenon" (this sounds as if a situation physically cannot be understood). The results instead are in-line, though weakly, with expectations from thermal fluctuations. It therefore appears the observations made with $\rho_{El}(T)$ are simply, at least partially, the physical consequence of the strong temperature dependency of the order parameter (and the weak temperature dependence of the normal conduction resistivity).

## 5    Consequence for unique identification of $T_{Crit}$

The boundary layer, $\delta T_\rho$, is not a materials layer but, in the $\rho_{El}(T)$-diagram, a small temperature interval within which resistivity changes drastically. This region can be interpreted as a temperature uncertainty $\delta T_\rho$ that exist *throughout* the superconductor material; the layer does not separate superconducting from normal conducting, *locally* different *materials* regions but indicates the transition interval between two different thermodynamic phases within the *same* superconductor material. The curve within this layer steadily approaches the convergence limit.

This is a counter-analogue to *local* temperature boundary layers, $\delta T$, for example in radiative transfer (mixtures of fine powders with strongly different extinction properties), or from boundary layers, $\delta v$, in fluid dynamics (insoluble mixtures of fluids with strongly different viscosity).

The *non-local* resistivity layer $\delta T_\rho$ is not associated with any solid boundary nor does it belong to a specific materials section. In the present picture, existence of $\delta T_\rho$ in the end results from the phase change that is quantified by the temperature dependency of the order parameter (its approximation in Eq. 2c) in a *homogenous* material the temperature of which is not uniform.

It is thus within the temperature interval, $\delta T_\rho$, in Figure 6b that $T_{Crit}$ cannot be defined sharply.

Besides conclusions drawn from solely the relaxation model (from the results condensed in Figure 11 in [25], part B), on whether $T_{Crit}$ can be defined uniquely in view of strongly diverging relaxation time, this is the second argument to question not only its sharp *definition* but even the mere *existence* of a uniquely defined, critical temperature in superconductors.

From Figure 6a,b, width of the boundary layer is estimated to between 11 (NbTi) and 20 (YBaCuO and BSCCO) µK, for decay (within $\delta T_\rho$) of $\rho_{El}$ by 1.68 e-6, 1.11e-5 and 1.22e-5 Ω m, respectively (these values of course depend on the estimated width). It is better to consider gradients: On the average, we have $\Delta\rho_{El}/\delta T_\rho$ = 0.153, 0.557 and 0.610 Ω m/K for NbTi, YBaCuO and BSCCO, respectively. The gradients (or their reverse values) allow estimates of the uncertainty by which critical temperature can be determined from resistivity (transport) measurements in these materials.

Reduction of the width $\delta T_\rho$ would be interesting to isolate (encircle) $T_{Crit}$ with reduced, finally perhaps vanishing uncertainty. A reduction of $\delta T_\rho$ might be found on the basis of correlations between $T_{Crit}$, $J_{Crit}$ and relaxation. Such correlations do exist, compare Figure 5a,b in [28]. But it is presently not clear whether a correlation between these critical parameters *and* the uncertainty $\delta T_\rho$ could be demonstrated.

## 6   Integration of relaxation into stability calculations

A general problem arises if the superconductor cannot relax to thermodynamic equilibrium within reasonable experimental or computation (simulation) time.

In reality, this is a problem of all standard (and up to now, also of the reported [4 - 6] numerical), stability calculations. The basic question behind reads: Is superconductor stability against quench just a matter of conductor geometry (diameter of fibres, thickness of thin films, aspect ratios), resistance, thermal diffusivity, critical and transport current distribution etc? In short, is it reduced to solely *engineering* aspects like they are considered in standard stability calculations? How can we obtain deeper understanding of the physics behind?

The "physics behind" is a problem of the dynamic behaviour of the *total* (emphasis is on "total") electron body and, after disturbances, its relaxation, in total, to a new dynamic equilibrium. The key to solve this problem is to calculate relaxation time as a function of the order parameter.

This problem becomes obvious when temperature is calculated either analytically or by means of numerical solutions of Fourier's differential equation. Superconductor temperature has carefully to be calculated, because, as we have seen in Figures 13b, and 14 of [7], the order parameter (as approximated by Eq. (2c)) depends very strongly on superconductor temperature.

In one dimension, Fourier's equation reads

$$\rho \, c_p([T(x,t)]) \, dT(x,t)/dt = \mathbf{div}[\lambda \partial T(x,t)/\partial x)] + Q(x,t) \qquad (8)$$

with local heat source, $Q(x,t)$, resulting from a disturbance, at a simulated process time, t. In Eq. (8), the operator "**div**" means divergence of heat flux, and $c_p$ and $\lambda$ are the specific heat and thermal conductivity of the material at positions (x,t), respectively.

The problem is twofold, items (i) and (ii):
Item (i): The commercially available, Finite Element code used in the transient temperature field calculations [4 - 6], like other FE-codes, does not differentiate between electron and lattice contributions to specific heat. Little information (except for some superconductor elements) is available for $c_{p,NC,}$ and $c_{p,SC,}$ the specific heat in the normal and superconducting state, respectively, at the same temperature. While unspecified values of $c_p$ are frequently reported in the literature when describing calculations of temperature excursions, electron and critical temperature, like the other superconductor critical parameters ($J_{Crit}$, $B_{Crit}$), refer to solely the *electron system* of the superconductor (though electrons and phonons are strongly coupled), which means, at $T < T_{Crit}$, the value $c_{p,SC}$ is needed (like from the dashed curve in Figure 4.1 of [13]).

It is not clear that $c_p$ in Eq. (8) could be taken just from $c_{p,SC} = c_p(T < T_{Crit})$ (like from the open circles in Figure 4.1 of [13]). When else using the dashed-dotted curve in this Figure (the much smaller $c_p$ of the lattice) as input into any FE code or simply into analytical calculations, too large values of T(x,t) would be obtained. Standard procedure with FE codes is to apply the $c_p$ of the lattice or an effective specific heat; this means their *standard* application does not (and cannot) provide temperature excursion of the electron system. Fourier's differential equation, i. e. calculation of temperature excursion in electrically neutral, continuum or dispersed systems, with existing temperature gradients, does not apply.

Item (ii) of the above: All calculated temperature excursions, T(x,t), and in particular $T_{Crit}$ are uniquely defined only if they are thermodynamic *equilibrium* values. Finite Element (FE) codes do not (and cannot) check whether thermodynamic equilibrium states are obtained during simulations. This is not to be confused with numerical convergence.

Equilibrium temperature, $T(x,t_{Eq})$, of the *electron* system with $t_{Eq} > t'$ (using the simulated process time, t'), with FE calculations presently can be obtained, in first approximation, when using a correction, a shift Δt(t'), that depends on simulation time, t'; the obtained T(x,t') has to be used to calculate Δt(t') and the result, in this approximation, has to be added to t'. The Δt(t') should, however, be obtained at the temperature of the *electron* system, but this temperature is presently not available from the FE results. We therefore have provisionally applied

$$T[x,t_{Eq}(t')] = T[t' + Δt(t')](x) = T(x,t') \qquad (9)$$

with T(x,t') the standard FE result. A solution of this problem might be found if Δt(t') is calculated iteratively; but this needs more investigations.

In any case, $T(x,t_{Eq})$ by Eq. (8) at $t_{Eq}$ is defined and is obtained only if relaxation from a disturbance is really completed and this state can successfully be integrated into the numerical procedure.

Each single relaxation step, of a series initialized at a temperature, $T(x,t)$, requests for its completion a very small time interval, $\partial t(t')$, and all the $\partial t(t')$ (a very large number) have to be summed up to the total $\Delta t(t')$.

Values of $\Delta t(t')$ have up to now been reported only for *filaments* [7]. In the following, we turn to *thin films*. In the conductor geometry of windings 96 to 100 shown in [8], Figure 1, thin superconductor films constitute a cable, a coated, multi-layer YBaCuO 123, thin film superconductor.

It is not clear that the results for filaments obtained in [7] or for thin films should be identical, first because of the different $T(x,t')$ in these systems that lead to very different $\Delta t(t')$ and $t_{Eq}$, also to different critical current density, $J_{Crit}[T(x,t')]$, and finally to different stability functions. The stability function is defined in the Appendix, Eq. (10a,b), to predict under which conditions zero-loss current transport is possible.

The $\Delta t(t')$ within the temperature region $T(x,t) \ll T_{Crit}$ are tiny, mostly below $10^{-9}$ ms, but if a disturbance starts very close to $T_{Crit}$, the shift becomes substantial, see later, Table 1. This is in accordance with the $\Delta t(t')$ previously reported for filaments.

For the present, *thin film* situation, to illustrate the extreme case resulting with the exponent n = 0.5 in the standard expression $J_{Crit}[T(x,t)] = [1 - T(x,t)/T_{Crit}]^n$, the temperature, e. g. $T(x,t') = 89.239$ K (at x = centroid coordinate of turn 96) is reached, as predicted by the Finite Element calculations, at t' = 4.200 ms after start of the simulations (the readers are kindly asked for a moment to tolerate large numbers of decimal digits; this is simply to demonstrate the shift is tiny provided temperature is definitely below $T_{Crit}$).

The thermal disturbance (local temperature increase by conduction of heat from other positions) initiated at this temperature and at this position causes a large number of electron pairs to decay. Their relaxation to a new, local equilibrium with electron pairs, from application of [7], takes Δt(t') = 1.422 e-8 ms so that equilibrium temperature $T(x,t_{Eq})$ = 89.239 K at the *same* position in reality is obtained not at t' = 4.200 ms but only later, i. e. not before $t_{Eq}$ = (4.200 + 1.422 e-8) ms, definitely a tiny correction.

Also at most of the other simulation times, t', the correction of the t' to the corresponding equilibrium values, $t_{Eq}$, is tiny. But the situation may change strongly if temperature increases more closely to $T_{Crit}$.

Under the same disturbance, with the same exponent n, at the same position (increase of centroid temperature solely by conduction), we have at T(x,t') = 91.933 K obtained from the FE calculations a Δt(t') = 1.191 e-3 ms and $t_{Eq}$ = 4.201 ms (the open red circles in Figure 7a,b). Further increase to T(x,t) = 91.9975 K yields even drastically increased Δt(t') and equilibrium $t_{Eq}$ (not shown in Figure 7b, but compare Table 1).

From technical and manufacturing aspects, and because of their improved current limiting properties, second generation (2G) thin film, coated YBaCuO 123 superconductors presently are considered more attractive than (1G) multi-filamentary BSCCO 2223/Ag tapes. Yet, when tentatively assuming the BSCCO 2223 superconductor material would be used for preparation of also the thin films of turns 1 to 100 in the same coil, with exactly the same conductor geometry, heat transfer to coolant and meshing, but simply with thermal diffusivity, critical current density and critical temperature of the BSCCO thin film material, Figure 8a,b shows that in comparison to Figure 5c of [8] conductor temperature increases significantly, and the "hot" turn now is turn 97 instead of turn 96 in case of the YBaCuO 123 thin film superconductor.

If in Figure 8b critical current density, $J_{Crit}$, decreases, and if, as is assumed in this Figure, voltage is constant, transport current density may become larger than (reduced) critical current density so that flux flow losses will increase. Increasing flux flow losses lead to increase of local conductor temperature, and this is confirmed in Figure 8b: The smaller $J_{Crit}$, the larger the local temperature, in Figure 8b the element temperature of the centroid.

Convergence of the numerical results in Figure 8a,b again is safely achieved and is almost perfect.

Calculations of temperature distributions, $f_S$, relaxation time, τ, from the shift, Δt(t'), at this temperature in both thin film superconductors takes enormous computational efforts. Even the numerical data input, before the proper calculations are started at all, takes about 15 min, which is explained by the large number of elements (in total about 65000) and the solid, thermal and electrical parameters assigned for each of the elements, with in total 10 different materials (like superconductor, stabilizer, solder, interfacial layers, metallic coating (Ag), buffer, substrate, casting compound, insulations). One simulation, in small time steps (between $10^{-14}$ and $10^{-6}$ s) extended over the period 0 ≤ t' ≤ 5 ms, requests more than 24 hrs computation time on a standard, 4-core PC under Windows 7.

It is clear that numerical simulations delicately depend on input parameters, details of conductor geometry (very different thicknesses and aspect ratios of very different materials layers involved, an issue that Finite Element calculations do not like at all), and on convergence schemes. Yet the results provide exact and time-resolved energy balances, temperature, critical current and transport current distributions that cannot be obtained with traditional stability models.

## 7   Open problems

Each of the simulation times, t', has to be understood as indicating *individual (initial) start points* (local disturbances, here at the centroid, **x**) caused by local events (absorption of a heat pulse, flux flow losses or simply a local temperature increase resulting from a disturbance at another, arbitrary positions, **x**', within the whole conductor cross section).

But the series t' and $t_{Eq}$ might be intermixed, which means the overall structure of T(x,t) would be lost, a value T(x,t'') might become smaller at a time t'' > t', which is impossible. This situation request more investigations. But with $T(x,t_{Eq}(x)]$ the local equilibrium temperature, resulting from a disturbance 1 occurring at a position **x'** that is overlaid onto the impact received from a disturbance 2 occurring at another position, **x''**, which means, after a different time shift, the totally resulting equilibrium temperature, given at the common $t_{Eq}$ that reflects the interfering $t_{Eq,1}(x')$ and $t_{Eq,2}(x'')$, with appropriate book-keeping in principle could be obtained provided a method can be found that does not rely on solutions of Fourier's differential equation in its original, basic form (calculation of temperature excursion in systems with existing temperature gradients).

As a consequence, the sequence of T(x,t') in total, i. e. the whole set of curves, not only the centroid curves in Figures 7a-c and 8b, but all curves at other positions, are shifted to *later* times by the transformations of the t' to the $t_{Eq}$ by Eq (7) while simulation temperature remains constant. The T(x,$t_{eq}$) thus are obtained as a set of equilibrium values, that are realized at the later times, $t_{Eq}$ > t'.

Since the Δt(t') are different for each T(x,t'), the *whole* set of equilibrium temperatures, T(x,$t_{Eq}$), for any position **x**, is distorted against T(x,t') which results in also distorted critical current density and distorted stability curves. This has been demonstrated in Figures 13a,b and 14 of [7] (these Figures are reprinted for comparison, Figure 9a,b). Note that contrary to Figures 7a-c and 8a,b of the present paper (thin films), they show the results for filaments. Because of some simplifying assumptions (arithmetic mean of Δt(t') taken instead of individual values, compare Figure Captions of Figure 9b), the distorted curves still resemble the original ones, but this similarity gets lost in Figures 7a-b and 8b when strictly the specific Δt(t') are taken to calculate $t_{Eq}$(t') = t' + Δt(t') for the plot of T(x,$t_{Eq}$).

From the relaxation time given by the term Δt(t') in Eq. (9), which means by the extremely large number of summations of individual ∂t(t'), it is clear the equilibrium temperature and the distortion of T(x,$t_{Eq}$) against T(x,t') must depend on the ratio ξ, of the active part of the electrons within the total electron body. Relaxation time, Δt(t'), is the larger, the larger ξ and thus the larger the superconductor simulation temperature, $T(x,t')$, and as a consequence, the smaller the order parameter.

The flat ellipse in Figure 7c collects all those $T(x,t_{Eq})$ that result from $T(x,t')$ and that are increasingly close to $T_{Crit}$. It shows that the $t_{Eq}$, as they may result from different t' (provided the corresponding $T(x,t')$ might be approximately equal and close to $T_{Crit}$) are intermixed, in the present case, chaotically; it is not uniquely clear that the $T(x,t_{Eq})$ can be ordered consistently. Without appropriate, complete book-keeping, the only conclusion that presently can be made is qualitative: The larger $T(t,')$, the larger are the shift $\Delta t(t')$ and the larger are the $t_{Eq}$. But identical $T(x,t_{Eq})$ might result from different events taking place at different t'.

The curves in Figure 9a,b, in contrast to Sect. 4 of [28], and to Figures 4b, 5a, 7, 9a,b and 11c of the same reference, result from single, isolated heat pulses ($Q = 3 \ 10^{-8}$ Ws and $2.5 \ 10^{-10}$ Ws, respectively, length 8 ns) applied to the conductor filament cross section, a less complicated, but strongly different situation in comparison to the events leading to the results in Figures 7a-c and 8a,b. "Less complicated" means: not by complex events like local flux flow losses or other individual events. In contrast, absorption of a single heat pulse, with its magnitude and position of its impact uniquely defined, is a comparatively simple event, at least in the sense of numerical simulations.

As has been shown in [8, 27] and is again demonstrated in Figures 7a-c and 8b, random, local, statistical variations of $T_{Crit}$ and $J_{Crit}$ (besides flux flow losses the "more complicated situation") are already sufficient to induce, without *any* fault current, just with transport current *constant* (!) and equal to nominal current, non-uniform temperature distributions.

In turn, without variations of $T_{Crit}$, $J_{Crit}$ and $B_{Crit,}$ and if density of transport current is below critical current density (no flux flow losses), the temperature distributions would be *flat* and would not show any temperature run-away to catastrophic divergence.

As before [7], the predicted distortion is more significant for the NbTi superconductor in comparison to the YBaCuO 123 and BSCCO 2223 thin films.

## 8    Summary

- For a clear understanding of how to avoid quench, the physics behind stability has to be analysed on a microscopic level.

- Stability analysis and predictions of superconductor stability may be of limited value if they do not consider superconductor relaxation after disturbances.

- Taking into account superconductor relaxation into stability calculations is mandatory if temperature of the electron system has already increased to values near critical temperature. This is shown for the thin film situation (the present paper) and confirms previous findings of the author reported for NbTi and for YBaCuO 123 filaments.

- If superconductor temperature, $T(x,t')$ approaches critical temperature very closely, distribution of the calculated equilibrium temperatures on the time axis presently may become chaotic: the $t_{Eq}$ cannot be ordered if there are several disturbances occurring in parallel, a situation that can be expected from e. g. local flux flow losses.

- Superconductor order parameter is the parenthesis that holds together (i) solution of the stability problem (unique identification of the temperature excursion on time axis), and as side results, (ii) bending of the specific resistivity, (iii) presently to only some extent, the impact of the thermal fluctuations problem, both at temperature near $T_{Crit}$. The order parameter thus is a very indicative quantity predicting the excursion with time of the superconductor under disturbances.

- Curvature of the resistivity at temperature below and above $T_{Crit}$ suggests definition of a non-local, resistivity "transition boundary layer" that may exist without any solid interface but is distributed throughout the superconductor volume. it is expressed as a temperature uncertainty interval, $\delta T_\rho$, around critical temperature. From the curves $\rho_{El}(T)$, gradients in the order of 0.1 to 1 ($\Omega$ m/K) can be extracted that allow to roughly estimate the uncertainty, at any position x, by which critical temperature can be determined from resistivity measurements of these materials.

- From the same resistances model, critical temperature, if it is understood as a *sharply, with zero tolerance defined* physical (thermodynamic) quantity, then would be a fiction. But exactly this follows from (a) the microscopic stability model (by the strong divergence of relaxation time) and (b) from the finite, non-zero temperature interval $\delta T_\rho$.

- A still open question concerns how temperature of both electron system and, in parallel, the lattice, could be achieved from numerical simulations, in order to apply Eq. (9) for calculation of $T(x,t_{Eq})$ at equilibrium time from the simulated $T(x,t')$ without iterations.

- All conclusions presented in this paper result from properties of many-particle systems (as previously explained, an analogy to nuclear physics), thermodynamic considerations (temperature uniquely defined under solely thermal equilibrium) and from an analogue to standard, multi-component heat transfer principles (solid conduction plus radiation in thin films).

**Data Availability Statement**

No Data associated in the manuscript.

Readers interested to reproduce the reported results should be familiar with, and have permission (valid licenses) to, apply standard, commercially available Finite Element (FE) codes and, if so, request dataset input files from the author. Instead of FE codes, Finite Differences (FD) codes, generated by their own development (like the FD code used in a first paper by the author to this subject), could be used as well by the readers, with increased computational efforts, however. Note that FE or FD codes are just *tools*, not the purpose, for investigation of the stability vs. relaxation problem.

# Appendix Limitations of the cell model and of calculation of total transport properties

**Current flow as a transport process**

Transport processes in general are described by Boltzmann's transport equation.

As a methodical tool, we in the following assume an electrical circuit that besides standard (Ohmic) resistances contains also *non-Ohmic* components like Josephson currents (Andreev reflections might be taken as another example but it is not very clear that this contribution arises in parallel to Ohmic components). This circuit shall serve as a model to check whether, or under which conditions, current flow can be described as a transport process.

Transport processes are initialized under electrical potential gradients (in thermal physics, under temperature gradients). Pure Ohmic current thus may be modelled as a current *transport* process across single or several Ohmic resistances. Andreev reflections, too, contribute to total current, but like Josephson currents cannot be simulated as a proper transport process: They do not need electrical potential differences for their existence.

For simulation of the Ohmic current contribution to total current, we apply the experimental, electrical resistivity, $\rho_{El}(T)$, from Figure VII - 11 in [24]. The values, $\rho_{El}(T)$, of YBaCuO 123 in Figure 1 of the present paper have schematically been extrapolated linearly from the region $T > T_{Crit}$ to $T_{Crit} - 2\ K \leq T \leq T_{Crit}$, see the green solid line in this Figure (an analogue would apply to also the other superconductor materials shown in the original

Figure VII - 11 in [24]). Strong deviations from the linear relationship would be expected only at $T \ll T_{Crit}$.

The extrapolated $\rho_{El}(T)$ serve for definition of a residual resistivity in the superconducting state if there were *no electron* pairs (like under very large magnetic field). In the cell model described in Subsect. 3.1 of the present paper, the resistivity $\rho_{El,SC}(T) = 0$ (or near zero) is obtained if pseudo-porosity $\Pi(T) \ll 1$ or if we have the ideal case (no down-bending, but the expected standard, sharp decrease of $\rho_{El,SC}(T)$ to finally zero at $T < T_{Crit}$). In both cases, the ideal value $\rho_{El,SC}(T) = 0$ results solely from contribution of the electron pairs to conductivity in that all normal conducting components are short-switched.

**A critique initiated by a parallel to multi-component heat transfer**
The following is a discussion of the question whether complex electrical or thermal transport processes can be described in terms of potential differences or temperature gradients, respectively. The readers if not very interested in details of, or parallels to, multi-component, conduction heat transfer, for clarification may go directly to [25, 26] (there with correspondingly calculated examples).

Using the extrapolated $\rho_{El}(T)$ in Figure 1, results obtained from the cell model for the total resistances of SC/NC/SC and SC/NC/NC contacts accordingly comprise only those electrons (of the active part) that contribute by Ohmic, not by exotic non-Ohmic processes like Josephson currents, to total current. This is because extrapolation of $\rho_{El}(T)$ is made from regions $T \gg T_{Crit}$ ($T > T_{Crit} + \delta T_\rho$, where only *normal* Ohmic conduction components of current exist), to the region $T < T_{Crit} - \delta T_\rho$.

The delicate point is: Determination of $T_{Crit}$ is done, in countless experiments, by measurement of the course of *total* resistance or resistivity within small intervals *around* $T_{Crit}$, but the mechanism of current transport in the two neighbouring intervals $T < T_{Crit} - \delta T_\rho$ and $T > T_{Crit} + \delta T_\rho$ is different. The proper residual resistance within $T \ll T_{Crit}$ accordingly must be different from values extrapolated from $T > T_{Crit}$ to $T \ll T_{Crit}$.

This situation is similar to multi-component heat transfer. If besides solid conduction e. g. radiation contributes to heat flow, the thermal conductivity or diffusivity (if it can be defined at all, see below) then apparently might depend on sample thickness (this is the heat transfer "thickness-effect" that was discussed in the 1980s literature, but has been solved meanwhile [25, 26]).

Unique definition of temperature and of $T_{Crit}$ is important for stability calculations and predictions since temperature fields obtained from FE calculations and their excursion with time are mapped onto the field of critical current density that in turn specifies the stability functions needed for stability predictions (see previous papers of the author). Only unique definition of temperature results in unique values of critical current density, $J_{Crit}$, and since the stability function, $\Phi$, contains only integrations over $J_{Crit}(x,y,t)$ dA (dA a cross section differential), calculation of $\Phi$ urgently needs clearly defined temperature fields with precisely known (as far as possible) individual values, $T(x,t)$. The fields $T(x,t)$ otherwise cannot suitably be mapped onto the fields $J_{Crit}(x,t)$ that are strongly temperature-dependent to yield the stability function,

$$0 \leq \Phi(t) = 1 - \int J_{Crit}(x,y,t) \, dA / \int J_{Crit}(x,y,t_0) \, dA \leq 1 \qquad (10a)$$

or its approximation

$$0 \leq \Phi(t) = 1 - \Sigma J_{Crit}(x,y,t)\, dA / \Sigma J_{Crit}(x,y,t_0)\, dA \leq 1 \qquad (10b)$$

In multi-component heat transfer, in order to define conductivity as a true materials property, the sample must be non-transparent to radiation. A sample is non-transparent to (direct) propagation of radiation if its optical thickness is large. Radiative and, as a consequence, total thermal conductivity *exist* only in this case.

"Existence" means: Conductivity can be specified as a true materials property, without any dependence on experimental parameters like sample thickness or thermal emissivity of enclosures that house materials samples. Only in this case can solid conductive and radiative components of multi-component heat transfer be calculated as independent of each other (as if the other component is not present at all), and only then can they be separated by temperature variations.

The question thus is under which conditions an effective, total electrical resistivity, $\rho_{El,SC}(T)$, or its reverse, an effective, total electrical conductivity, as a genuine materials property, in analogy to the thermal analogue, would exist in exotic cases with non-Ohmic components in general, and whether each component can be calculated "as if the other is not present" and, by analogy, whether the components can be separated experimentally, by variations of voltage.

The question therefore is: What is the analogue (if there is any) to radiative non-transparency that could facilitate calculation of total electrical current and separation into independent contributions in case a circuit contains also non-Ohmic components?

In multi-component heat transfer, large optical thickness, τ → ∞, transforms the very complex integro-differential equation (the "Equation of Radiative Transfer") to a 2nd order differential equation, like Fourier's differential equation. The so called "Additive Approximation" [25], which means algebraic addition of conductivities, becomes applicable under solely *this* condition.

What then is the analogue to large optical thickness that would

(i) get electrical resistivity a genuine, solely *materials* property in that it does not depend on experimental parameters?
(ii) allow conductivity components to be calculated independent of each other, experimentally be separated and algebraically be added to a total conductivity?

In heat transfer calculations, summation of conductive and radiative components to total heat flux $q_{Total} = q_{Cond} + q_{Rad}$ (Ws/m$^2$), beyond doubt is correct, clearly from phenomenological viewpoints. But it is, in general, not correct to simply add corresponding conductivity components, $\lambda_{Cond}$, $\lambda_{Rad}$ (W/(m K)), to a total, solid plus radiative conductivity, $\lambda_{Total} = \lambda_{Rad} + \lambda_{Cond}$. This is a thermally "exotic" example of energy transport (but is realistic, because it can come up in thin films). The components might be coupled by the temperature profile in a sample, which means they would not be independent of each other or even would not exist at all (in the meaning of "existence" explained above).

In this then not clear that a temperature gradient and thus a radiative conductivity would exist at all positions within a sample including its boundaries. In multi-component heat transfer, also a *total* conductivity therefore exists only (and total heat flux can be calculated from a temperature gradient taken over the total sample thickness) if a temperature gradient exists, *everywhere* within the sample, which means: if temperature profile, T(x), is differentiable in x, at all positions.

By analogy, the simple, algebraic addition of electrical current components, $I_k$, in Eq. (11) (Eq. 6-9 of [13]), of course is correct, like the summation $q_{Total} = q_{Cond} + q_{Rad}$ in thermal physics is correct, again from the phenomenological viewpoint. The point is that the authors [13], p. 320, say, below their Figure 6.8: "This means in the end that $I_q$ (note by the present author: $I_q$ the quasi-particle current) is treated in terms of an Ohmic resistance R." Their Eq. (6-9) in [13] yields

$$I = I_j + I_q + I_v \qquad (11)$$

using $I_q = U/R$..

The parallel to heat transfer is obvious. Total electrical current, I, could be calculated only if all components in Eq. (11) could be assigned electrical potential gradients.

While quasi-particles resemble real particles quite closely, "the real particle plus as cloud of agitated particles (virtual energy states), constitutes the quasi-particle (the cloud screens the real particle, which means a quasi-particle interacts only weakly with other quasi-particles." (this citation, set in quotation marks, is from R. D. Mattuck, [29], p.14).
Quasi-particles thus are not locally specified, materials bodies, they rather are *correlations* between real particles and excitations, like electron pairs are better described by correlations between single, real particles, each of elementary charge, e, not by materials bodies of charge 2e. The distance between electrons i and j of a pair is not uniquely defined, which means consideration an electron pair as a solid particle of charge 2e is doubtful.

The question thus remains whether it is possible to explain a current *without* recourse to potential differences and "resistances"? If so, are *all* the components in Eq. (6-9) of [13] *not* switched exactly in parallel*,* but approximately in series, to each other? Is it not possible to experimentally to separate the components?

In multi-component heat transfer, this is quite different since radiative conductivity is obtained as the result of a differential process: The optical thickness is extended to infinity, and the radiative conductivity converges to an expression based on just the *wall* temperature of a container that houses the non-transparent sample (interestingly not on *internal* sample temperatures).

In Eq. (11), the component $I_q$ is explained as current across a resistance. But if the other components cannot be explained by resistances, can they really be separated from $I_q$?

In summary, items 1 and 2:
(1) Total, multi-component heat transfer (heat flow or heat flux) cannot be calculated from temperature gradients and conductivity or diffusivity values if any component of the energy equation cannot be written in terms of a conduction process, i. e. by a differential expression.
(2) Total, multi-component current cannot be calculated from electrical potential gradients if any component of total current is not of Ohmic type.

The cell model explained in this paper thus is applicable only in case there are no non-Ohmic contributions to total resistance and total current.

# Figures

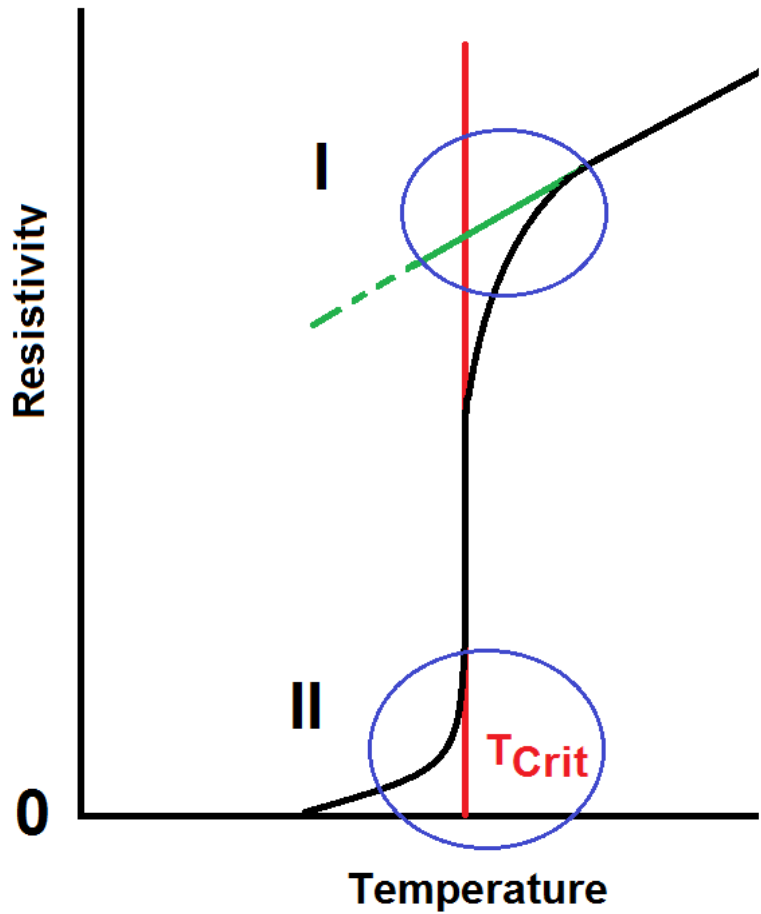

Figure 1 Resistivity, $\rho_{El}(T)$, of superconductors near critical temperature (schematic). For examples see e. g. Figure VII - 11 in the book "Copper Oxide Superconductors" by Poole, Datta and Farach [24] that shows bending of the resistivity vs. temperature curves of YBaCuO 123 and of other superconductor compounds in which a Rare Earth element has been substituted for the element Y in the $REBaCuO_{7-x}$ family. At $T_{Crit}$, the coloured, open ellipses I and II highlight well-known deviations from the sharp increase of $\rho_{El}(T)$ usually expected during a warm-up (in this schematic Figure, the deviations are exaggerated).

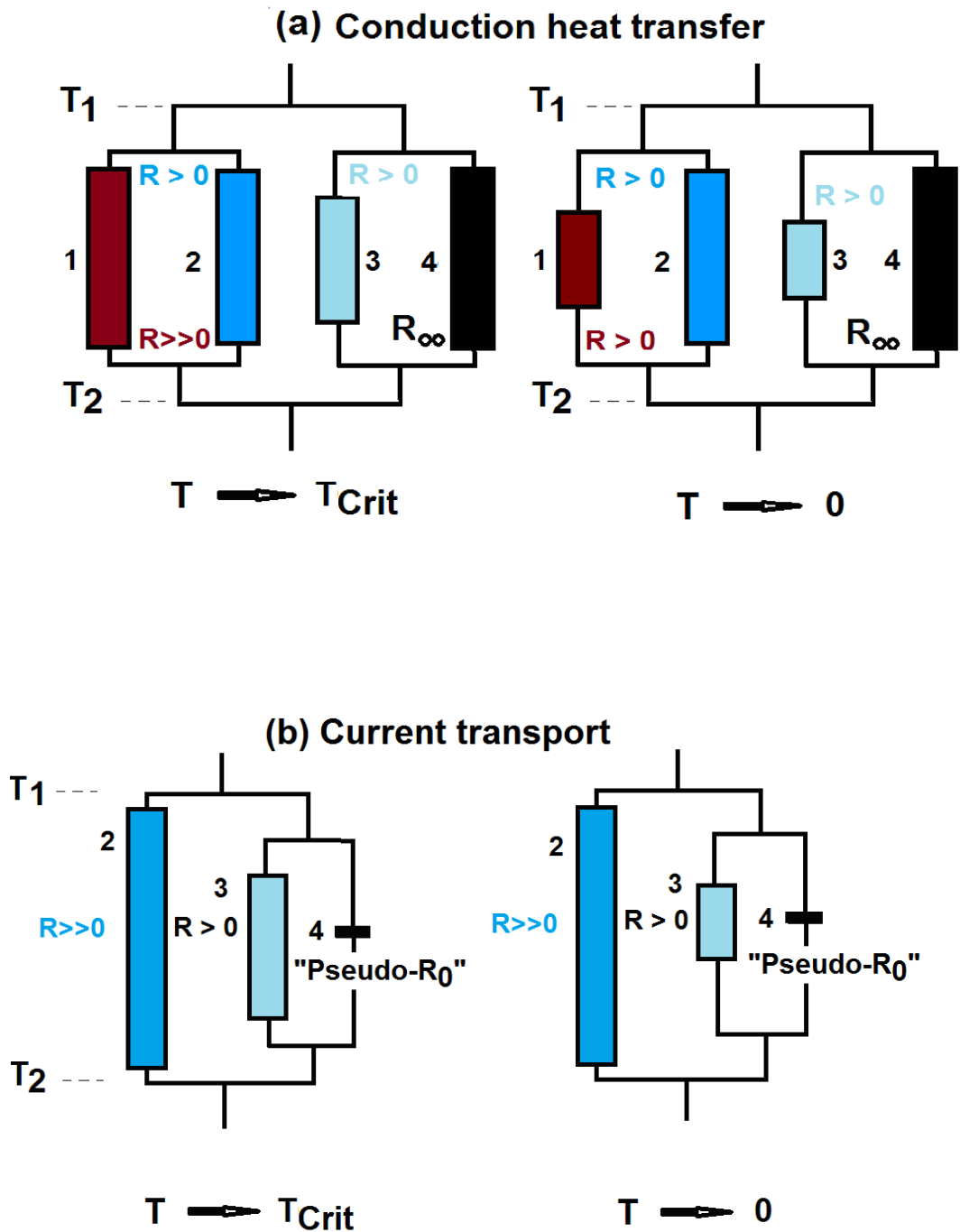

Figure 2a,b Resistance networks: (a) thermal and (b) electrical resistances, all at T < $T_{Crit}$ of a superconductor. The diagrams are used to specify application of the Russell cell model [19] for thermal and electrical conduction transport in the material (schematic, strongly simplified). The percentage of electrons (of the *total* body) that are available ("active") for thermal and electrical transport applies to the composite of resistances 3 (light-blue) and 4 (black) in both diagrams (a) and (b). The dark-blue rectangles denote the single (not condensed to pairs) electron contribution to resistance of the *residual* body, and the brown rectangles denote the thermal resistance by the lattice (phonons).

(a) *Conduction heat transfer* at different temperature. Dark brown and dark-blue thermal rectangles indicate non-zero phonon and single electron *thermal* resistances, (1) $R_{Ph}$, (2) and (3) $R_{El}$, respectively. The black resistance (4) $R_\infty$, applies to electron pairs and illustrates their infinitely large *thermal* resistance (regardless of their number). Their zero contribution to conduction heat transfer results from vanishing collisions with the lattice. Vertical length of the black rectangles schematically indicates increasing number of electron pairs. At very low temperature, heat transfer in both (a) and (b) being subject to resistances $R_{Ph} > R_{El}$, the thermal conductivity of the superconductor to the most part is by single (not condensed) electrons. Under given temperatures, $T_1$ and $T_2$, standard solution of Fourier's differential equation yields the phonon temperature within resistance 1 that, at any co-ordinate, is different from electron temperature in resistances 2 and 3. Temperature of channel 3 like channels 1 and 2 is below $T_{Crit}$ but otherwise undetermined.

(b) *Electrical current transport*, like in (a) at different temperature (again schematic, strongly simplified). Against (a), dark-brown rectangles (1) are cancelled (totally insulating, electrical transport channel). Single (not condensed) electrons contribute, according to non-zero, electrical resistances, $R_{Ph}$ and $R_{El}$, respectively. The black rectangles (resistances $R_o$) illustrate zero resistance of electron pairs. Regardless of their number, total contribution to current transport is by electron pairs only. In order to make the Russell cell model (as a conduction or resistance model) applicable for the simulations, an at least 20 orders of magnitude smaller electrical resistance, in relation to normal electrical conduction, has to be assumed for the superconductor (this is schematically indicated by the small, non-zero vertical length of the black rectangles).

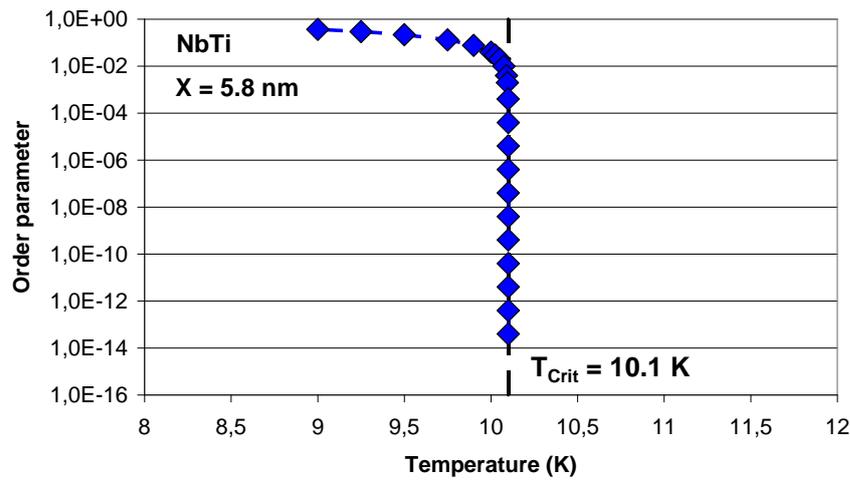

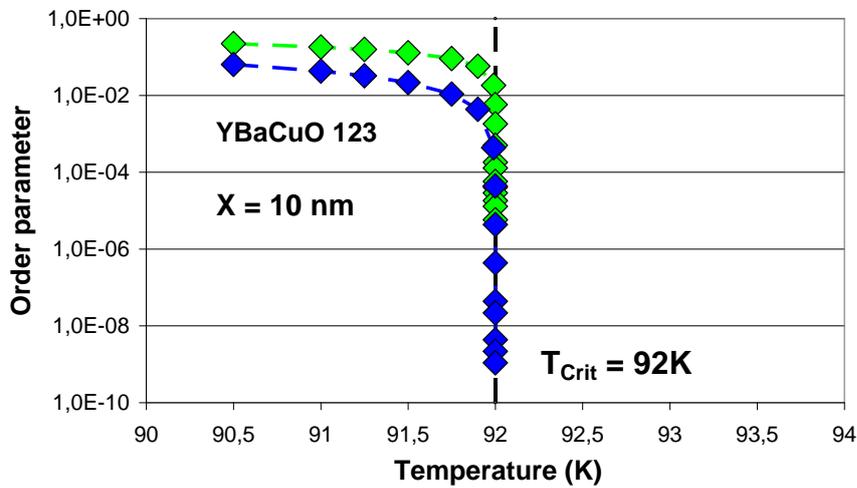

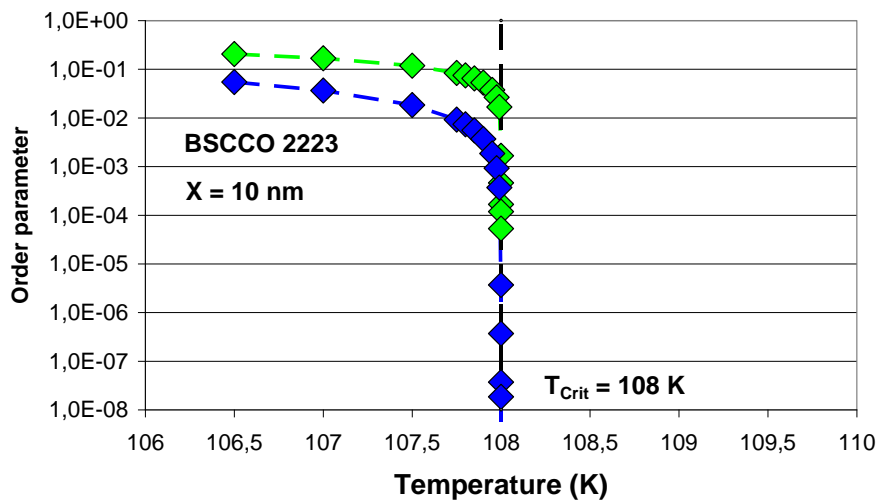

Figure 3 Ratio $f_S = n_S(T)/n_S(T=4K)$ (an approximation to the proper, Ginzburg-Landau order parameter, compare Eq. 2a-c), of NbTi, YBaCuO 123 and BSCCO 2223 vs. temperature. Results are obtained using either the microstability (relaxation) model

[7] or the approximation by Eq. (8) of [12] (light-green and dark-blue symbols, respectively). Electron pair density within the *active* electron part at T = 4 K in thermodynamic equilibrium is 3e26/m$^3$.

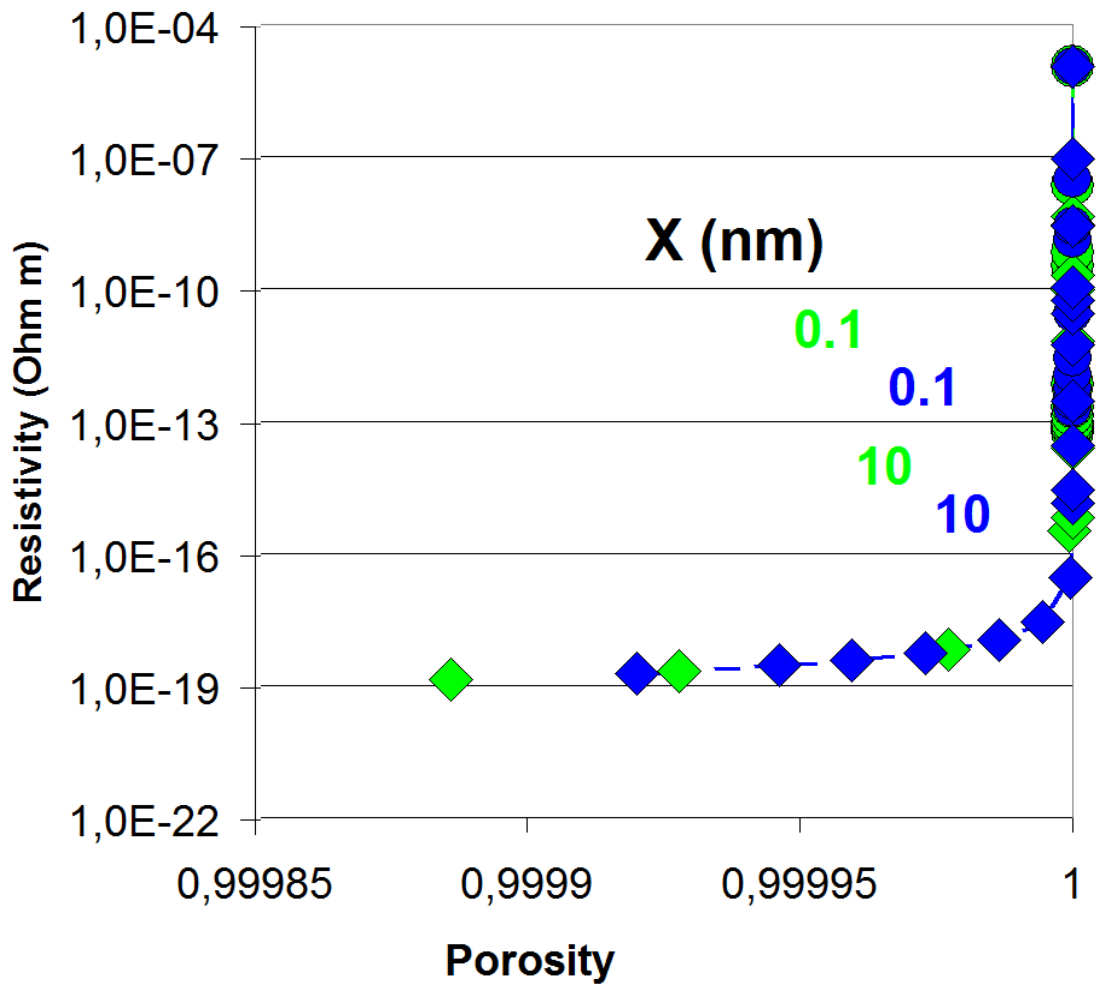

<u>Figure 4</u> Dependence of the resistivity, $\rho_{El,SC}$, on pseudo-porosity of YBaCuO 123 using different coherence lengths, X (light-green and blue symbols). The calculations, to obtain the ratio $f_S$ (Figure 3), porosity, Π, and resistivity, $\rho_{El,SC}$, apply the microscopic (relaxation) stability model ([7], solid diamonds) or the approximation Eq. (8) of [12] (solid circles), respectively. Under increasing temperature, all curves, Π,) and $\rho_{El}(T)$, steadily converge to Π = 1 and to $\rho_{El,NC}$ (the normal conduction value), respectively. The sharp increase of the resistivity is due to the results found for the order parameter (in its approximation, Eq. 2a-c).

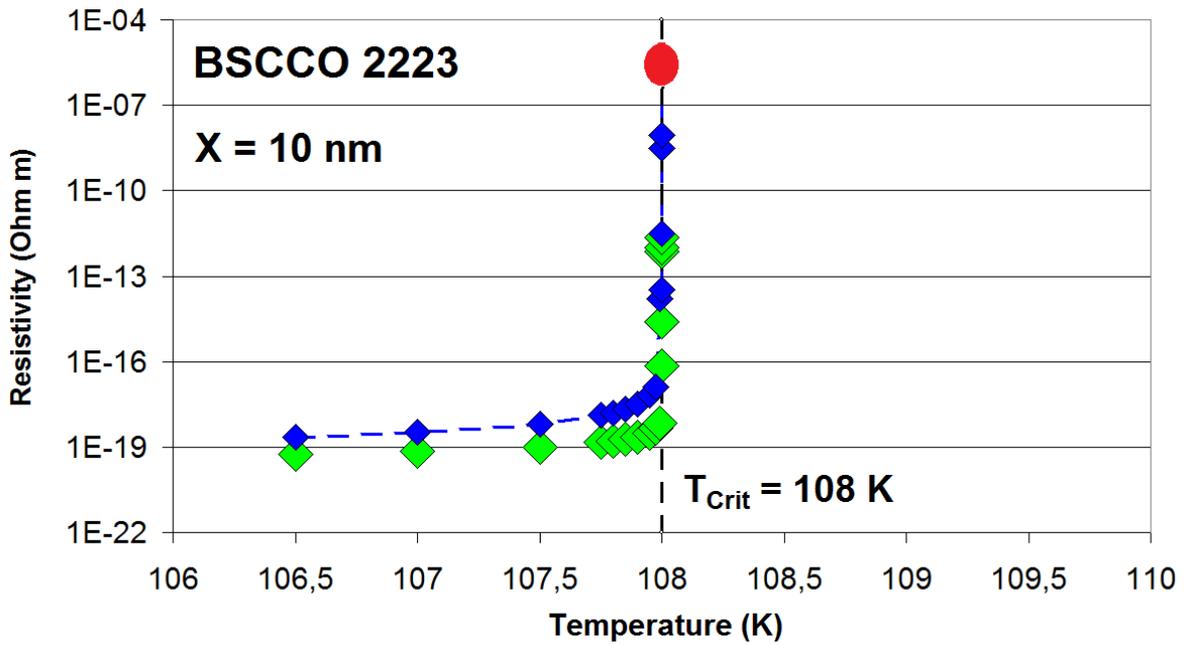

Figure 5a Effective electrical resistivity, $\rho_{El}(T)$, of BSCCO 2223 near thermal phase transition. The results are obtained *before* point-symmetry operations (see text) have been performed. The curves are obtained using the Russell cell model [19], with the ratio, $f_S(T)$, that approximates the order parameter, and porosity, $\Pi$, either from application of the microscopic stability model [7] (light-green) or from application of an approximation (Eq. (8) in [12], blue, solid diamonds). The value of the superconductor pseudo-conductivity, $\rho_{el,SC}$, amounts to $10^{-25}$ $\Omega$ m. Curvature of $\rho_{El}(T)$ below $T_{Crit}$ is confirmed, for any variations of $\rho_{El,SC}$ between this value and $10^{-50}$ $\Omega$ m. Meaning of the red solid circle is explained in Caption to Figure 5b. Note the logarithmic scale of the resistivity axis.

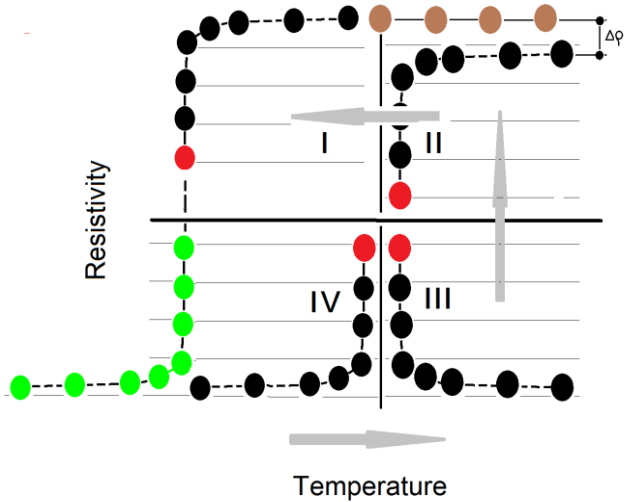

Figure 5b Stepwise explanation of point symmetry (mapping) of the $\rho_{El,SC} = f(T)$, here indicated by solid, black circles (schematic). The symmetry operation starts in quadrant IV with the $\rho_{El,SC}$ shown in Figure 5a. The red, solid circle denotes the maximum numerical value obtained in the series $\rho_{El,SC} = f(T)$, in this quadrant. Mapping is continued (light-grey arrows) counter-clock wise to the final result seen in quadrant I (the real values are shown in Figure 6a). The light-brown, solid circles denote the normal conduction resistivity, $\rho_{El,NC}$, of the superconductor. See text for more explanations.

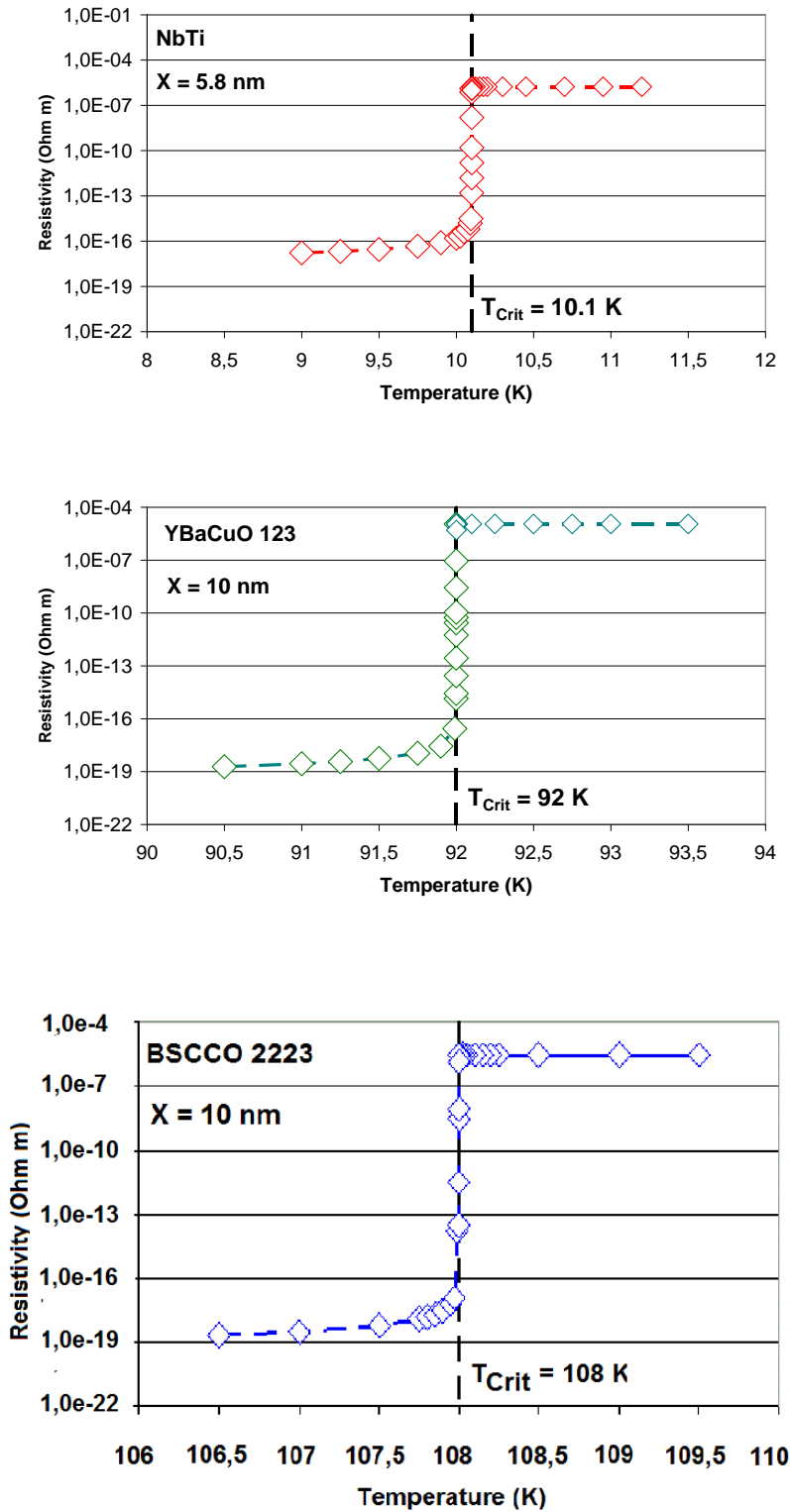

Figure 6a Resistivity vs. temperature, the "full curves" obtained *after* completion of the point-symmetry operation (extended to quadrant I) onto the results calculated for the NbTi, YBaCuO 123 and BSCCO 2223 superconductors. While in the region T < $T_{Crit}$, between $10^{-19}$ and $10^{-13}$ Ω m, bending of the curves near 108 K is clearly seen, it cannot be revolved to the same extent between $10^{-7}$ and $10^{-4}$ Ω m because of the logarithmic resistivity scale. Open symbols are applied to improve visibility of the curvatures under logarithmic resistivity scales; these are seen more clearly if the conductivity, instead of the resistivity, is considered in Figure 6c.

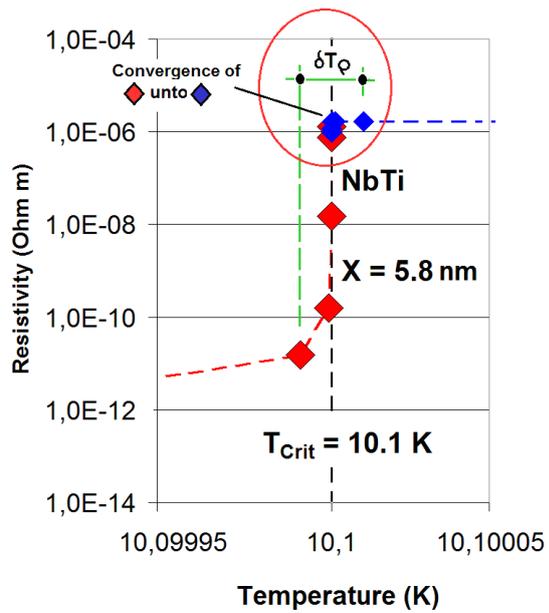

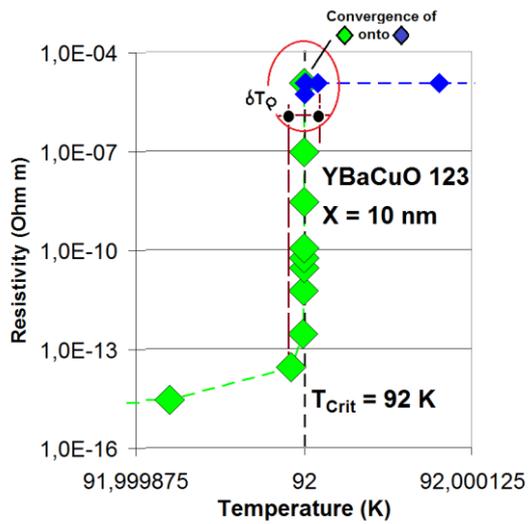

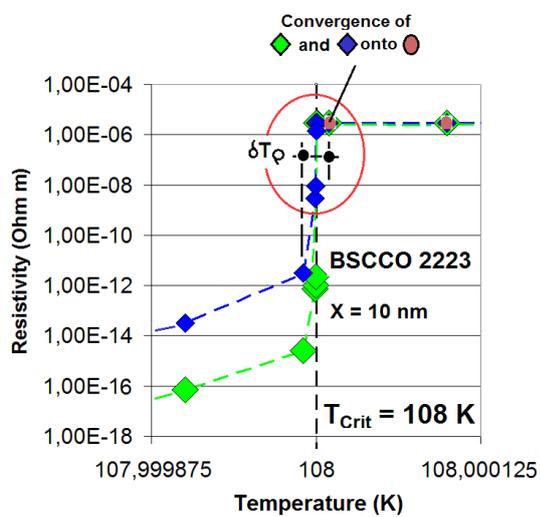

<u>Figure 6b</u> Details of the resistivity vs. temperature curves of NbTi, YBaCuO 123 and BSCCO 2223 obtained after completion of the point-symmetry operations extended to quadrant I (see text for explanation). The horizontal temperature scale is strongly magnified. All diamonds converge to the normal conduction resistivity, $\rho_{El,NC}$, at $T > T_{Crit}$ (the light-brown, solid circles in the bottom diagram) of the superconductor. The Figure shows, very close to $T_{Crit}$, the non-local, resistivity boundary layer expressed as the temperature uncertainty, $\delta T_\rho$, in the three materials.

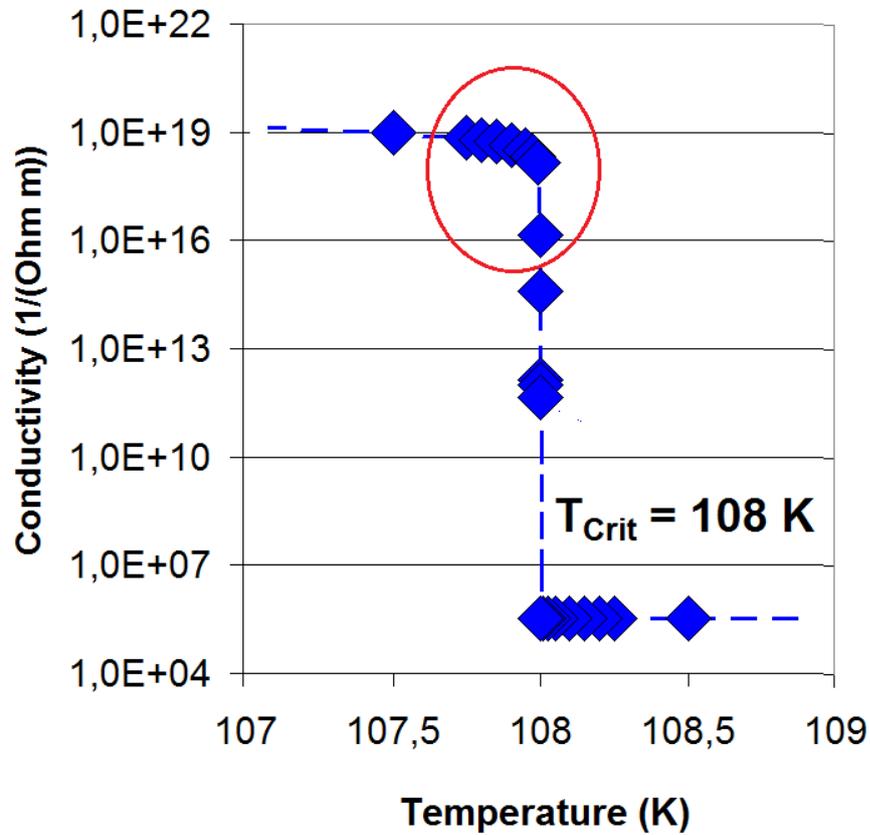

Figure 6c Electrical conductivity, $1/\rho_{El,SC}$, of BSCCO 2223 calculated from the resistivity values shown in Figure 6a. The open, red circle highlights bending of the curve that here is more clearly seen than in Figure 6a,b.

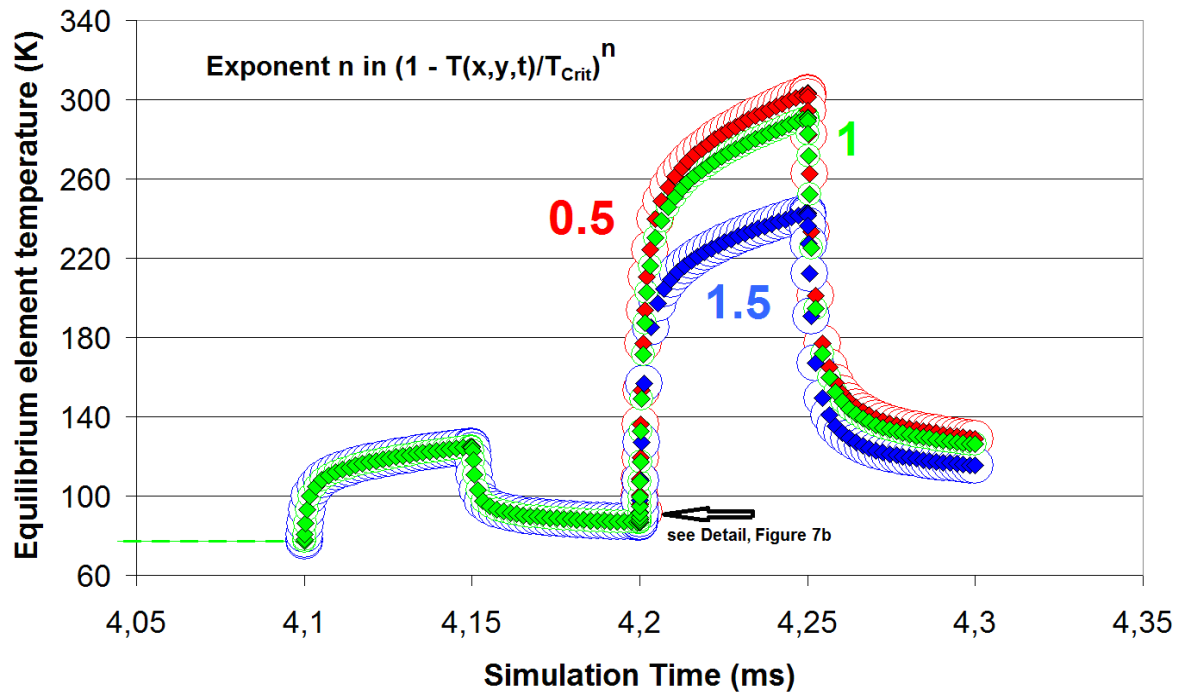

Figure 7a Equilibrium element temperature, T(x,t) of the centroid in turn 96 (for the cable geometry, compare Figure 1 in [8]). Results are given for different values n of the exponent in $J_{Crit}[T(x,t)] = [1 - T(x,t)/T_{Crit}]^n$ for the temperature dependency of critical current density (a standard relation for $J_{Crit}$ (T), with n = 1.5 the Ginzburg-Landau exponent). Solid symbols show temperature in dependence of simulation time, t', open circles show the same temperature but vs. equilibrium time, $t_{Eq}$. Compare text, Sect. 6, for explanation of the difference between both time scales. The almost hidden, open red circle (compare the black arrow) indicates $t_{Eq}$ that at T = 91.9325 K differs from t' only very slightly (see Table 1, n = 0.5). The corresponding $t_{Eq}$ at T = 91.9975 K that is substantially larger (in the order of 100 ms) than its t' is not shown. The Figure is copied from its original (Figure 5a in [28], here without the "convergence circles".

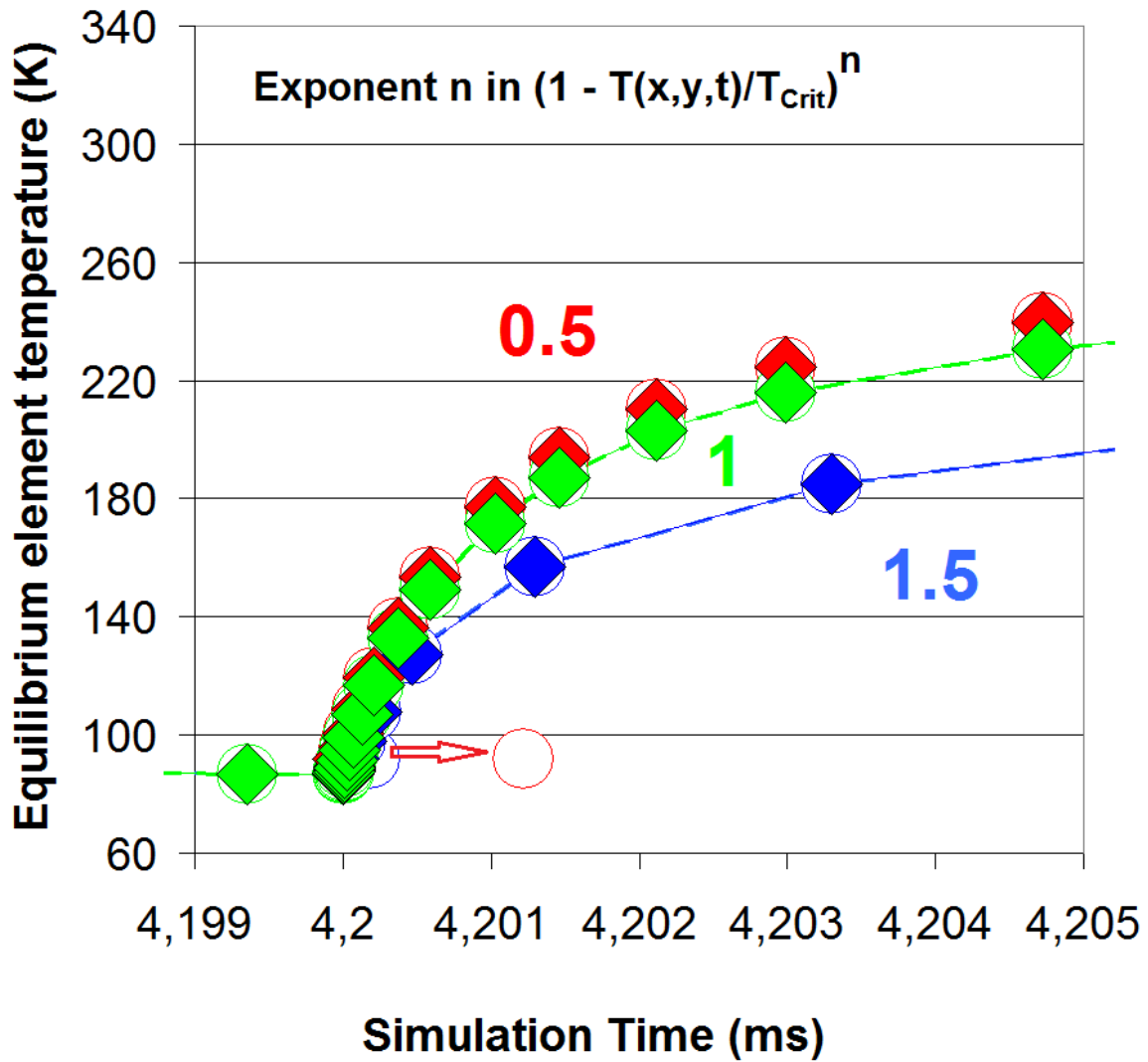

Figure 7b Same plot as in Figure 7a but in detail showing the deviation of $t_{Eq}$ from t' of a data point obtained from the FE calculations near phase transition. The open red circle indicating the corresponding equilibrium temperature is shifted from t' to the definitely larger $t_{Eq}$. The shift of simulation time in T(x,t'), namely t' → $t_{Eq}$) to time scale, $t_{Eq}$, yielding T(x,$t_{Eq}$). is indicated by the red arrow (temperature, T, remains unchanged).

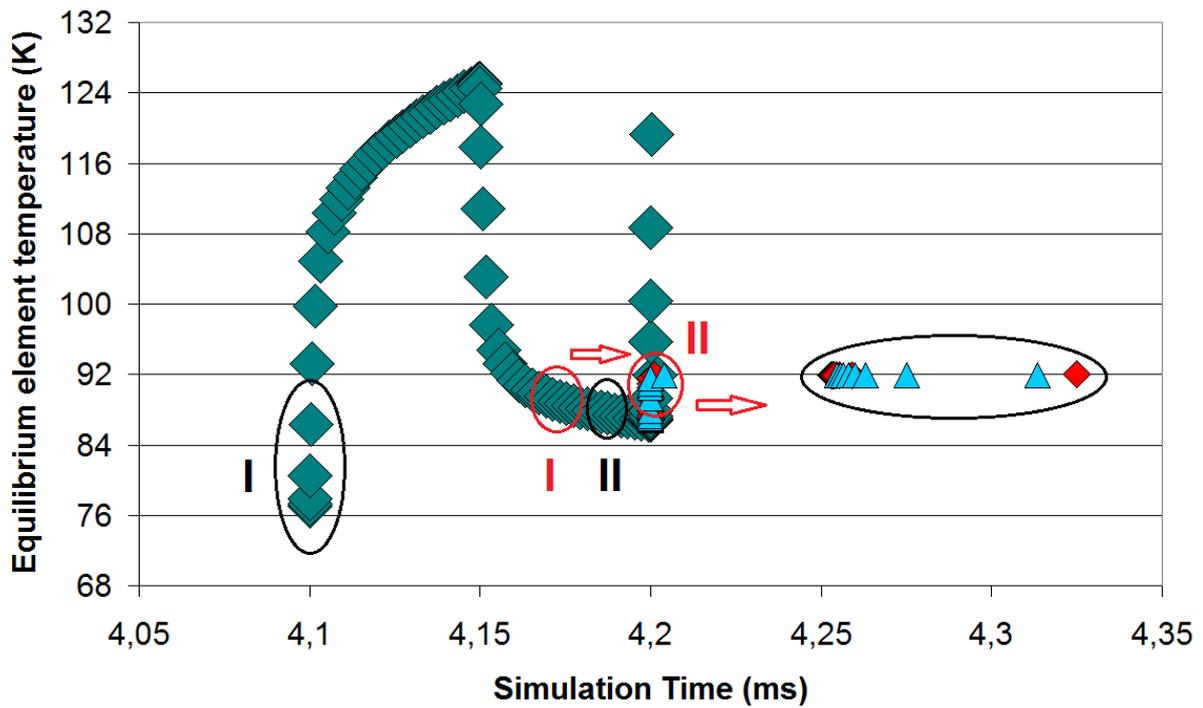

<u>Figure 7c</u> Same plot as in Figure 7a but in more detail showing the deviation of $t_{Eq}$ from t' of data points obtained from the FE calculations near phase transition. The more closely temperature approaches the phase transition, the larger is the shift (the bulged-out calculated curve), as resulting from relaxation, from simulation times, t' (dark green diamonds) to the corresponding equilibrium times, $t_{Eq}$ (data points in the flat ellipse). Results are given for the exponent n = 0.5 in $J_{Crit}[T(x,t)] = [1 - T(x,t)/T_{Crit}]^n$ and for different values of the ratio ξ, that indicates the "active" part of the electron body that contributes to thermal and current transport and specific heat. Red and dark-green diamonds and light-blue triangles correspond to ξ = 5, 10 and 15 percent, respectively. The larger ξ, the larger is the shift Δt(t'), and the larger is the relaxation time. As explained in the text, this linear transformation is provisional only (an exact calculation of $T(x,t_{Eq})$ by FE simulations or by iterations is presently not possible.

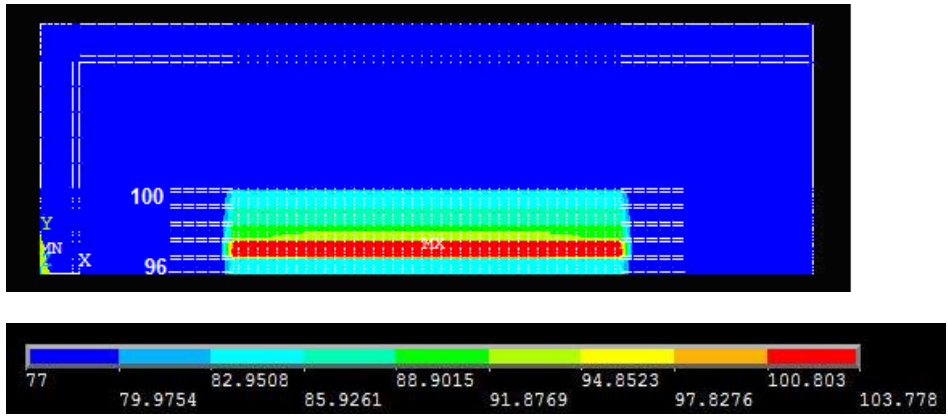

Figure 8a Equilibrium nodal temperature distribution within the cable cross section (turns 96 to 100). White dashed lines are part of the mesh (the inner block comprises turns 96 to 100 of the coil; the narrowly spaced double white lines indicate electrical insulation between turns, and the outer double lines reflect a reinforcement of the casting compound). This temperature distribution is obtained if instead of YBaCuO 123 the superconductor material BSCCO 2223 tentatively would be used to prepare thin films. All results obtained for $J_{Crit}$(77K) = 1.05 $10^9$ A/m$^2$ within 0 ≤ t' ≤ 4.2 ms safely converge during the numerical simulations, in this Figure for simulation time t' = 4.1 ms (note that coolant temperature, 77 K, is confirmed and the position "MX" of the temperature maximum is located at almost exactly the conductor mid position). In comparison to Figure 5c of [8], the YBaCuO 123 case, the maximum temperature is found in turn 97 instead of 96, and the distribution of conductor temperature is smooth, with no fluctuations, and is smaller in the thin film material (does not exceed $T_{Crit}$ = 108 K), which means flux flow losses may occur if local current transport density, $J_{Trans}$(x,y,t') is larger than critical current density.

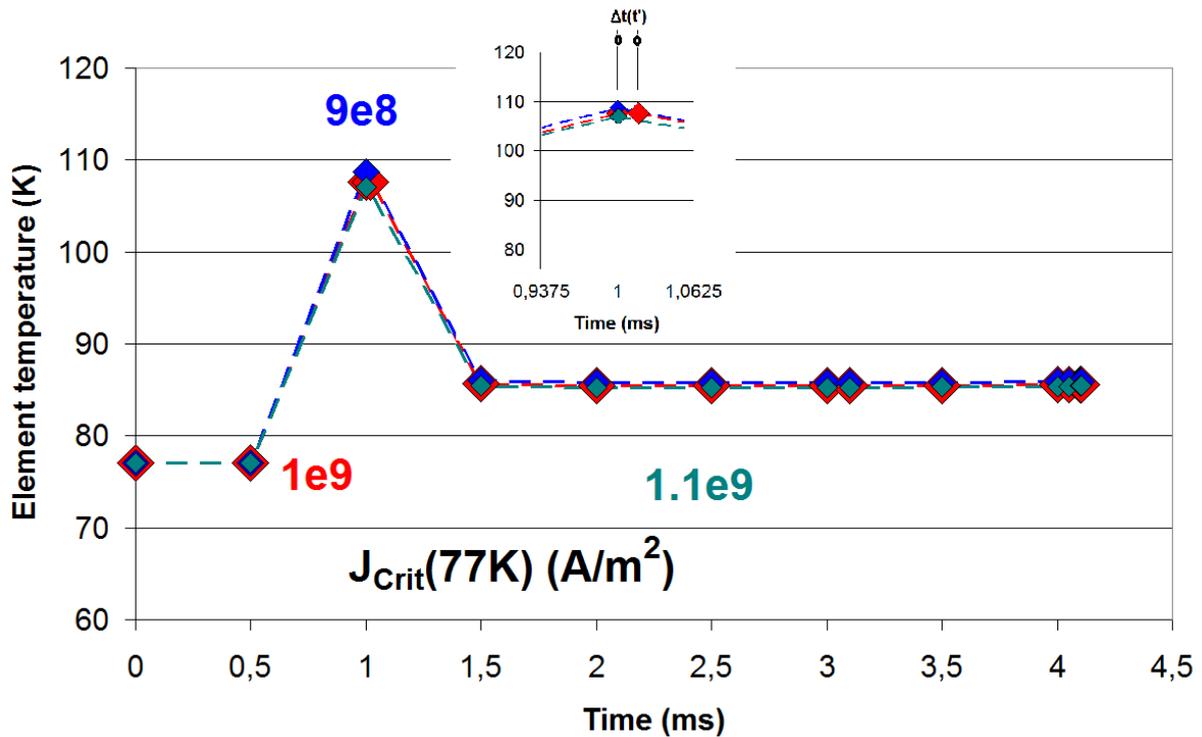

Figure 8b Equilibrium element temperature, T(x,t) of the centroid in turn 96 (same cable geometry as in Figure 7a-c). Numerical calculations are performed as in Figure 7a but now for application of the BSCCO 2223 superconductor material, within exactly the same conductor cross section and meshing. Results are given for the value n = 1.5 of the exponent in $J_{Crit}[T(x,t)] = [1 - T(x,t)/T_{Crit}]^n$ for the temperature dependency of critical current density, but for small variations (dark-blue and dark-green diamonds) of $J_{Crit}(77K)$ against the mean, $10^9$ A/m$^2$ (red diamonds, note the sequence of blue, red and green diamonds), for constant (maximum, minimum) voltage peaks. The inset (detail near simulation time t' = 1 ms) shows the shift Δt(t') = 1.7701 $10^{-5}$ ms resulting for $J_{Crit}(77K) = 10^9$ A/m$^2$. Compare again text, Sect. 6, for explanation.

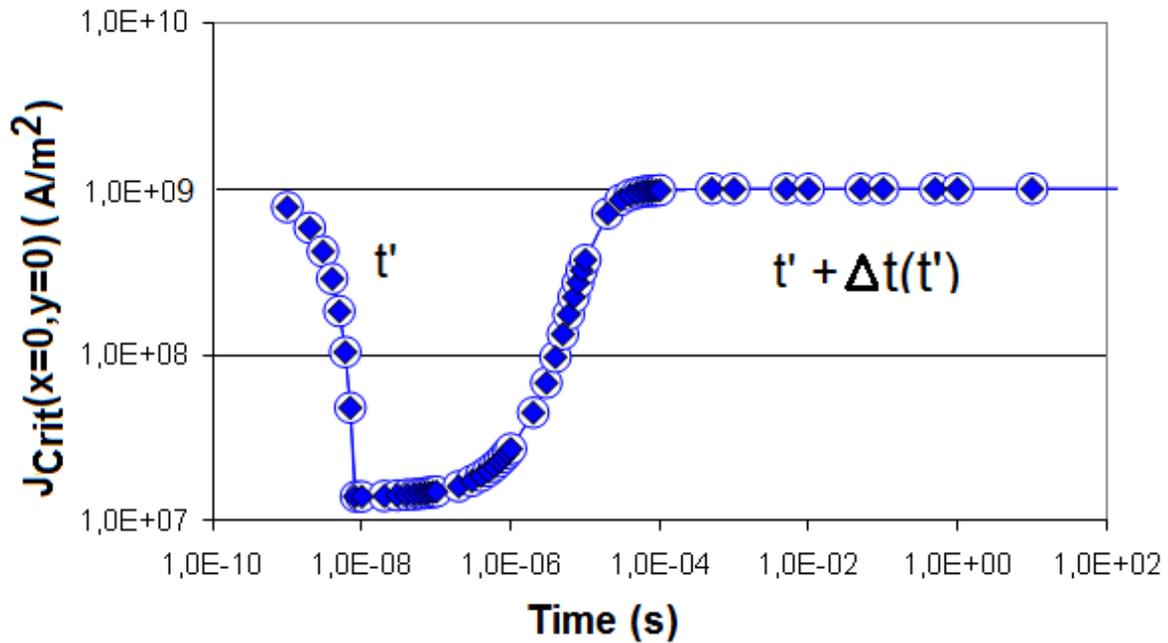

Fig. 9a Results reported in [7], reproduced here for comparison (and for convenience of the readers). Critical current density, $J_{Crit}(x,y,t)$, in the superconducting YBaCuO 123 filament cross section. Data are calculated from the element temperatures using the exponent n = 2 and are given for the element positioned near the central node (x = 0, y = 0). The $J_{Crit}$ (x,y,t) are plotted vs. real (i. e. simulation) time, t' (solid symbols) and the "shifted" time scale, t → $t_{Eq}$ = t' + Δt(t') (open symbols), with the shift Δt(t'). A pulse of Q = 3 $10^{-8}$ Ws is absorbed at radial positions during a period of 8 ns. See the original Figure Caption in Figure 13b of [7]. Conductor geometry is shown in Figure 1a of [5].

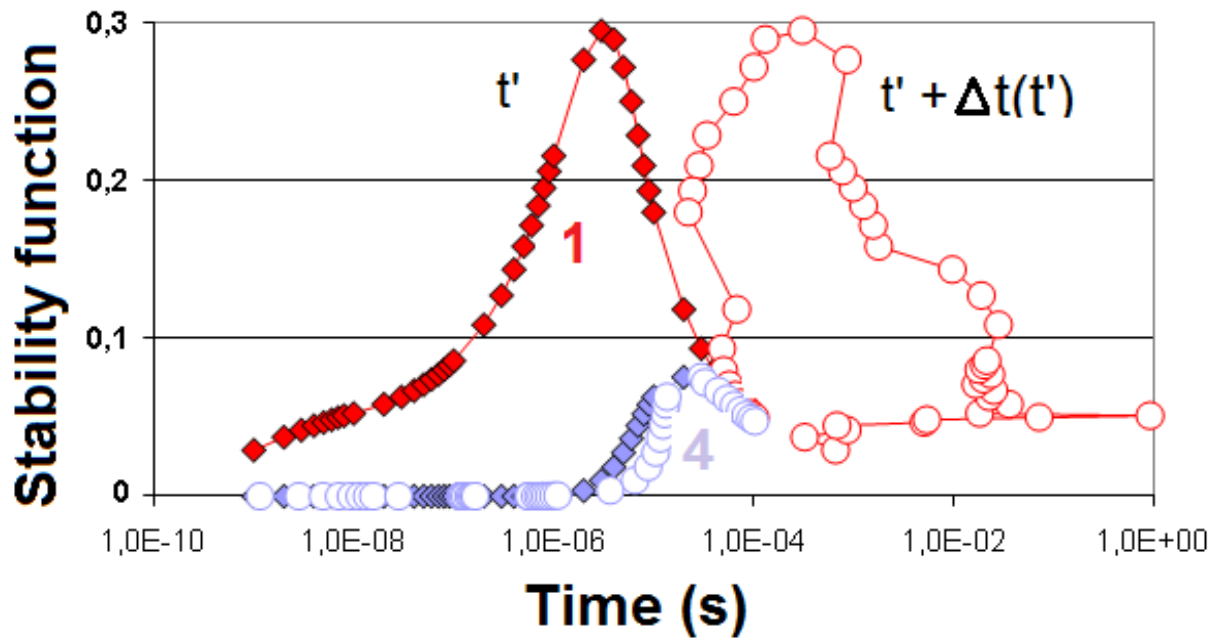

<u>Fig.9b</u> Results reported in [7], reproduced here for comparison. Stability function, Φ(t), of the NbTi-filament, calculated using a heat pulse absorbed at radial positions 0 ≤ x ≤ 6 µm, y = 0, of Q = 2.5 $10^{-10}$ Ws during a period of 8 ns. The figure shows Φ(t) at planes 1 and 4 (axial distances from the target spot of y = 0 and 56.3 µm, respectively). Compare Figure 1a in [5] for conductor geometry. Data Φ(t), copied from Figure 14 of [7], are plotted vs. real (i. e. simulation) time scale, t' (solid symbols) and the "shifted" time scale t → $t_{Eq}$ = t' + Δt(t') (open symbols), as a rough approximation with an arithmetic mean of the shift taken over the corresponding planes.

| T(x,t) of Centroid (K) | t' (ms) | Δt(t') (ms) | $t_{Eq}$ (ms) | In Figure 7a,b: |
|---|---|---|---|---|
| 89,23869312 | 4,200011800E+00 | 1,422111958E-08 | 4,200011814E+00 | |
| 91,93253103 | 4,200025400E+00 | 1,191142117E-03 | 4,201216542E+00 | red open circle |
| 91,9975 | 4,252099273E+00 | 9,862376318E+01 | 1,028758624E+02 | not shown |
| | | | | |
| 87,74630188 | 4,200005000E+00 | 4,246666079E-09 | 4,20000500424667E+00 | |
| 89,04192278 | 4,200011800E+00 | 1,165701175E-08 | 4,20001181165701E+00 | not shown |
| 91,5303986 | 4,200025400E+00 | 2,607124594E-06 | 4,20002800712459E+00 | |
| | | | | |
| 92,12951274 | 4,159348400E+00 | 0,000000000E+00 | 4,159348400E+00 | |
| 91,36549101 | 4,161348400E+00 | 1,062653419E-06 | 4,161349463E+00 | not shown |
| 90,77014004 | 4,163348400E+00 | 1,508428464E-07 | 4,163348551E+00 | |

<u>Table 1</u> Transformation of simulation time, t', to equilibrium time, $t_{Eq}$, by the shift Δt(t') (time needed for relaxation). Red, blue and light-green numbers result for n = 0.5, 1.0 and 1.5, respectively, in $J_{Crit}[T(x,t)] = [1 - T(x,t)/T_{Crit}]^n$. Results are given for the ratio ξ = 0.1 for the active electron part of the total electron body. Results for other ratios are given in Figure 7c. With n = 1.5, the Ginzburg-Landau exponent, the ratio ($t_{Eq}$ - t')/[$T_{Crit}$ - T(t')] becomes very small. The diverging number of digits shall just demonstrate the shift Δt(t') is tiny if temperature is clearly below $T_{Crit}$.